\documentclass[11pt,a4paper]{article}
\pdfoutput=1 % if your are submitting a pdflatex (i.e. if you have
\RequirePackage{ifpdf}
\usepackage{amsmath} 

\usepackage{jheppub}
\usepackage{pstricks}
\usepackage[final]{pdfpages}
\usepackage{ifpdf}
\usepackage{slashed}

\usepackage{hyperref}

\usepackage{color}
\usepackage{graphics}

\usepackage{etoolbox} % Load before 'breqn' package to make 'lineno' work

\title{Higgs Rapidity Distribution in $b {\bar b}$ Annihilation at Threshold in N$^{3}$LO QCD}

\author[a]{Taushif Ahmed,}
\author[a]{M.~K. Mandal,}
\author[a]{Narayan Rana}
\author[b]{and V. Ravindran}

\affiliation[a]{Regional Centre for Accelerator-based Particle Physics,\\ Harish-Chandra Research Institute, Allahabad, India}
\affiliation[b]{The Institute of Mathematical Sciences, Chennai, India }

\emailAdd{taushif@hri.res.in}
\emailAdd{mandal@hri.res.in}
\emailAdd{narayan@hri.res.in}
\emailAdd{ravindra@imsc.res.in}

\abstract{We present the rapidity distribution of the Higgs boson produced through bottom quark annihilation at third order in QCD using the threshold approximation. We provide a framework, based on the factorization properties of the QCD amplitudes along with Sudakov resummation and the renormalization group invariance, that allows one to perform the computation of the threshold corrections in a consistent, systematic and accurate way. The recent results on threshold N$^3$LO correction in QCD for the Drell-Yan production and on three loop QCD correction to Higgs form factor with bottom anti-bottom quark are used to achieve this task. We also demonstrate the numerical impact of these corrections at the LHC.} 
\preprint{HRI-RECAPP-2014-024}

\begin{document}
\allowdisplaybreaks[4]
\unitlength1cm
\maketitle
\flushbottom

%************
% Definition
%************

\def\D{{\cal D}}
\def\DD{\overline{\cal D}}
\def\g{\overline{\cal G}}
\def\gm{\gamma}
\def\M{{\cal M}}
\def\ep{\epsilon}
\def\unM{\hat{\cal M}}
\def\unas{ \left( \frac{\hat{a}_s}{\mu_0^{\epsilon}} S_{\epsilon} \right) }
\def\rnM{{\cal M}}
\def\rnas{ \left( a_s  \right) }
\def\bt{\beta}
\def\cD{{\cal D}}
\def\cC{{\cal C}}
\def\ca{\text{\tiny C}_\text{\tiny A}}
\def\cf{\text{\tiny C}_\text{\tiny F}}
\def\ct{{\red []}}

%*******
% Intro
%*******

\section{Introduction}
\setcounter{equation}{0}
\label{sec:intro}

With the spectacular discovery of the Higgs boson at CERN LHC \cite{1207.7214, 1207.7235}  , the full spectrum of matter particles and force carriers of the Standard Model (SM) has been established very successfully. Though the mass of the newly discovered boson is already pinned down with an impressive experimental uncertainty of just a few hundred MeV in the range of 125 - 126 GeV, to fully validate the mechanism of electroweak symmetry breaking and to shed light on the possible potential deviations from its SM apprehension, it is indispensable to study the inclusive as well as exclusive observables associated with the production and decay channels of the Higgs boson to a very high accuracy.

Within the framework of SM, the production mechanism of the Higgs boson is dominated by gluon fusion, whereas one of the alternative channels, namely, bottom quark annihilation is severely suppressed by the small Yukawa coupling of bottom quark to the Higgs boson. However, in extensions of the SM with an enlarged spectrum of Higgs sector, as in the case of two-Higgs doublet model, the Yukawa coupling of bottom quark to some of the Higgs bosons can be enhanced significantly, such that the production channel of bottom quark annihilation could be the dominant one. Moreover, the contribution from gluon fusion channel decreases due to enhanced negative top-bottom interference diagrams. Furthermore, the bottom quark initiated processes at hadron colliders are of much theoretical interest on account of the freedom in treating the initial state bottom-quarks. In the four flavor scheme (4FS), alternatively known as the fixed flavor number scheme (FFS), the mass of the bottom quarks is considered to be non-zero throughout and they are excluded from the proton constituents, whereas, in the framework of five flavor scheme (5FS), also known as the variable flavor number scheme (VFS), the bottom quarks are considered as massless partons, except in the Yukawa coupling, with their own parton distribution functions (PDF).

The inclusive productions of the Higgs boson in gluon-gluon fusion \cite{Georgi:1977gs, Djouadi:1991tka, Dawson:1990zj, Spira:1995rr, Catani:2001ic, Harlander:2001is, Harlander:2002wh, Anastasiou:2002yz, Ravindran:2003um}, vector boson fusion \cite{Bolzoni:2010xr} and associated production with vector bosons \cite{Han:1991ia} are known to next-to-next-to leading order (NNLO) accuracy in QCD. The Higgs boson production through bottom-antibottom ($b \bar b$) annihilation is also known to NNLO accuracy in the VFS \cite{Dicus:1988cx, Dicus:1998hs, Maltoni:2003pn, Olness:1987ep, Gunion:1986pe, Harlander:2003ai}, whereas it is known to NLO in the fixed FFS \cite{Reina:2001sf, Beenakker:2001rj, Dawson:2002tg, Beenakker:2002nc, Raitio:1978pt, Kunszt:1984ri}.

While the theoretical predictions at NNLO and next-to-next-to leading log (NNLL) \cite{Catani:2003zt} QCD corrections and of two loop electroweak effects \cite{Aglietti:2004nj, Actis:2008ug, Altarelli:1978id, Matsuura:1987wt, Matsuura:1988sm, Hamberg:1990np} played an important role in the discovery of the Higgs boson, the theoretical uncertainties resulting from the unphysical factorization and renormalization scales are not fully under control.  In addition, the interpretation of the experimental data with higher accuracy from the upcoming run at the LHC demands the inclusion of higher order terms in QCD in the theoretical computation. Hence, the efforts to go beyond NNLO are going on intensively in past few decades. The computation of N$^3$LO corrections is underway and some of the crucial ingredients, like the quark and gluon form factors \cite{Moch:2005id, Moch:2005tm, Gehrmann:2005pd, Baikov:2009bg, Gehrmann:2010ue}, the mass factorization kernels \cite{Moch:2004pa} and the renormalization constant \cite{Chetyrkin:1997un} for the effective operator describing the coupling between the Higgs boson and the SM fields in the infinite top quark mass limit are available up to three loop level in dimensional regularization. In addition, NNLO soft contributions are also known \cite{deFlorian:2012za} in $n$ dimensions.  These results were already used to compute the partial threshold contributions at N$^3$LO to the production cross-section of di-leptons in Drell-Yan (DY) and of the Higgs boson in gluon fusion as well as in $b\bar{b}$ annihilation, see \cite{Moch:2005ky, Laenen:2005uz, Idilbi:2005ni, Ravindran:2005vv, Ravindran:2006cg}. Since then, there have been several advances \cite{Kilgore:2013gba, Anastasiou:2013mca, Duhr:2014nda, Dulat:2014mda} towards obtaining the complete N$^3$LO result for the inclusive Higgs production. The milestone in this direction was achieved by Anastasiou et al. in \cite{Anastasiou:2014vaa} to obtain the complete threshold N$^3$LO corrections. This result provided a crucial input in \cite{Ahmed:2014cla} to obtain the corresponding N$^3$LO threshold corrections to DY production.  Independently, in \cite{Li:2014bfa}, using light-like Wilson lines threshold corrections to the Higgs boson as well as Drell-Yan productions up to N$^3$LO were obtained.  Catani et al. in \cite{Catani:2014uta} used the universality of soft gluon contributions near threshold and the results of \cite{Anastasiou:2014vaa} to obtain general expression of the hard-virtual coefficient relevant for N$^3$LO threshold as well as threshold resummation at next- to-next-to-next-to-leading-logarithmic (N$^3$LL) accuracy for the production cross section of a colourless heavy particle at hadron colliders. There have been several attempts to go beyond threshold corrections \cite{Presti:2014lqa, deFlorian:2014vta} for the inclusive Higgs production at N$^3$LO.  Recently, \cite{Anastasiou:2014lda}, the full next to soft as well as the exact results for the coefficients of the first three leading logarithms at this order have been obtained for the first time. For the Higgs boson production through $b \bar{b}$ annihilation, the recent results of the Higgs form factor with bottom-antibottom by Gehrmann and Kara \cite{Gehrmann:2014vha} and the universal soft distribution obtained for the Drell-Yan production \cite{Ahmed:2014cla} enabled us to obtain the missing $\delta(1-z)$ contribution (see \cite{Ravindran:2005vv, Ravindran:2006cg,Kidonakis:2007ww} for the partial results to this order) to the production cross-section at threshold at N$^3$LO \cite{Ahmed:2014cha}.  

Like the inclusive one, the differential rapidity distributions are computed for the dilepton pair in DY \cite{Anastasiou:2003yy} and the Higgs boson produced through gluon fusion in \cite{Anastasiou:2004xq, Anastasiou:2005qj}, the Higgs boson through $b\bar{b}$ annihilation in \cite{Buehler:2012cu} and associated production of the Higgs with vector boson in \cite{1107.1164, 1312.1669} to NNLO in QCD. Using the formalism developed in \cite{Ravindran:2005vv, Ravindran:2006cg}, the partial N$^3$LO threshold correction to the rapidity distributions of the dileptons in DY and the Higgs boson in gluon fusion as well as bottom quark annihilation were computed in \cite{Ravindran:2006bu}. Following the same technique, we obtained the complete N$^3$LO threshold correction to the rapidity distributions of both dilepton pair in DY and the Higgs boson in gluon fusion \cite{Ahmed:2014uya}. We had seen the dominance of the threshold contribution to the rapidity distribution in these processes. A significant amount of reduction in the dependence on the unphysical renormalization and factorization scale of the rapidity distribution takes place upon inclusion of the N$^{3}$LO threshold corrections. In addition, these computations provide first results beyond NNLO level and will serve as a non-trivial check for a complete N$^3$LO results. Keeping these motivations in mind, we intend to extend the existing result of the rapidity distribution of the Higgs boson produced through $b{\bar b}$ annihilation to higher accuracy, namely the inclusion of complete N$^3$LO threshold correction. 

In Sec.~\ref{ThresNLO}, we perform an explicit calculation of threshold correction to the rapidity distribution of the Higgs boson in $b{\bar b}$ annihilation at NLO, using the factorization properties of QCD amplitude, Sudakov resummation of soft gluons and renormalization group invariance. This helps us to build an elegant framework to calculate the rapidity distribution at threshold, of a colorless state produced at hadron colliders, to all orders in QCD perturbation theory. In Sec.~\ref{ThresBNLO}, we use that general framework to achieve the goal of computing the complete analytic expression for the threshold corrections beyond NLO and provide the result up to N$^{3}$LO. Sec.~\ref{NumRes} contains the discussion on the numerical impacts of our results. Finally, we conclude with our findings in Sec.~\ref{Conclude}.

%****************************************************
% Differential distribution with respect to rapidity
%****************************************************

\section{Differential Distribution with Respect to Rapidity}
\label{DiffRap}

The interaction of bottom quarks and the Higgs boson is encapsulated in the following action 
\begin{eqnarray}
S^b_{I} = - \frac{\lambda}{\sqrt{2}} \int d^4 x \, \phi(x) \overline \psi_b(x) \psi_b(x) 
\end{eqnarray}
where, $\psi_b(x)$ and $\phi(x)$ denote the bottom quark and scalar field, respectively. The Yukawa coupling $\lambda$ is given by $\sqrt{2} m_b/v$, with the bottom quark mass $m_b$ and the vacuum expectation value $v\approx 246$ GeV.  Throughout our calculation, we consider five active flavours (VFS scheme), hence except in the Yukawa coupling, $m_b$ is taken to be zero like other light quarks in the theory.  

We study infrared safe differential distribution, namely rapidity distribution of the Higgs boson at hadron colliders, in particular those produced through bottom anti-bottom annihilation. Our findings are very well suited for similar observables where the rapidity distribution is for any colorless state produced at hadron colliders. We will set up a framework that can provide threshold corrections to rapidity distribution of the Higgs boson to all orders in perturbation theory. It is then straightforward to obtain fixed order perturbative results in the threshold limit.

The general frame work that we set up for the computation of threshold corrections beyond leading order in the perturbation theory for such observables is based on the factorization property of the QCD amplitudes. Sudakov resummation of soft gluons, renormalization group equations and most importantly the infrared safety of the observable play important role in achieving this task. QCD amplitudes that contribute to hard scattering cross sections exhibit rich infra-red structure through cusp and collinear anomalous dimensions due to the factorization property of soft and collinear configurations. Massless gluons and light quarks are responsible for soft and collinear singularities in these amplitudes and also in partonic subprocesses. Singularities resulting from soft gluons cancel between virtual and real emission diagrams in infrared safe observables.  While the final state collinear singularities cancel among themselves if the summation over degenerate states are appropriately carried out in such observables, the initial state collinear singular configurations remain until they are absorbed into bare parton distribution functions. In the upcoming section, we present one loop computation for the rapidity distribution in order to demonstrate how the various soft singularities cancel and also to give a pedagogical derivation of how the most general  resummed threshold correction to the rapidity distribution can be obtained.

\subsection{Threshold Correction at NLO}
\label{ThresNLO}

The process under consideration is the production of the Higgs boson through bottom quark annihilation in hadron colliders. The leading order process is  
\begin{equation}
b (k_1) + \bar{b} (k_2) \rightarrow H (q)  
\end{equation}
where, $k_i$'s are the momenta of the incoming bottom and anti-bottom quarks involved in partonic reaction and $q$ is the momentum of the Higgs boson. The hadronic center of mass energy squared is defined by $S \equiv (p_1+p_2)^2$, where $p_i$'s are the hadronic momenta and the corresponding one for the incoming partons is given as $\hat{s} = (k_1+k_2)^2$.  The fraction of the initial state hadron momentum carried by the parton is denoted by $x_i$ \textit{i.e.} $k_i = x_i p_i$.
The rapidity of the Higgs boson is defined through
\begin{equation}
 y = \frac{1}{2} \ln \left( \frac{p_2 \cdot q}{p_1 \cdot q} \right) \, .
\end{equation}
The differential distribution with respect to rapidity of the Higgs boson can be expressed as 
\begin{equation}
\label{sigmaW}
\frac{d }{dy }\sigma^b(\tau,q^2,y) = 
\sigma^{b,(0)}(\tau,q^2,\mu_R^2) W^b(\tau,y,q^2,\mu_R^2) ~,
\quad \quad \quad \sigma^{b,(0)}= {\pi \over 4 S N}
 \lambda^2(\mu_R^2) 
\end{equation}
with $\tau \equiv {q^2}/{S}$, $q^2=m_H^2$, $m_{H}$-the mass of the Higgs boson. $\lambda(\mu_R^2)$ is the Yukawa coupling defined at the renormalization scale $\mu_R$, $N=3$ is the number of QCD colors and $\sigma^{b, (0)}$ is the leading order cross-section. Defining $z \equiv {q^2}/{\hat s}$, we find  
\begin{align}
 W^b(\tau,y,q^2,\mu_R^2) &= {\left(Z^b(\mu_R^2)\right)^2\over \sigma^{b,(0)}}
\sum_{a,c=b,\overline b,g} ~ \int_{0}^1 d x_1 \int_{0}^1 d x_2 ~ \hat {\cal H}_{ac}(x_1,x_2) \int_0^1 dz ~ \delta (\tau - z x_1 x_2)
 \nonumber\\
 & \times \int dPS_{1+X} ~ |{\cal M}_{a c \rightarrow H+X}|^2 ~ \delta \left(y - \frac{1}{2} \ln \left( \frac{p_2\cdot q}{p_1\cdot q} \right)\right) \,.
\end{align}
In this expression, $X$ is the remnants other than the Higgs boson, $Z^b(\mu_R^2)$ is the ultraviolet (UV) renormalization constant for the Yukawa coupling $\lambda$ and $dPS_{1+X}$ is the phase space element for the $H+X$ system. ${\cal M}_{a c\rightarrow H+X}$ denotes the scattering amplitude at partonic level.
The function $\hat {\cal H}_{ac}(x_1,x_2)$ is the product of unrenormalized parton distribution functions (PDF) $\hat f_a(x_1)$ and $\hat f_c(x_2)$, 
\begin{eqnarray}
\hat {\cal H}_{ac}(x_1,x_2)
\equiv\hat f_a(x_1) \hat f_c(x_2) \, .
\end{eqnarray}
The PDF $f_a(x_1,\mu_F^2)$, renormalized at the factorization scale $\mu_F$, is related to the unrenormalized ones through Altarelli-Parisi (AP) kernel $\Gamma_{ad}$ as follows:
 \begin{eqnarray}
f_a(x_i,\mu_F^2) = \sum_{d=b,\overline b,g}\int_{x_i}^1 {dz \over z}~ \Gamma_{ad}( \hat a_s,\mu^2,\mu_F^2, z, \epsilon)~ \hat f_d\left({x_i \over z}\right), \quad \quad a=b,\overline b,g 
\end{eqnarray}
where, the scale $\mu$ is introduced to keep the strong coupling constant  $\hat g_s$ dimensionless in space-time dimensions $n = 4 + \epsilon$, regulating the theory and $\hat a_s \equiv \hat g_s^2/16 \pi^2$. Expanding the AP kernel in powers of ${\hat a}_{s}$, we get
\begin{eqnarray}
 \Gamma_{ad}( \hat a_s, \mu^2,\mu_F^2, z, \epsilon)= \delta_{ad}\delta(1-z) + \hat a_s S_\epsilon \left({\mu_F^2 \over \mu^2}\right)^{{\epsilon\over 2}}
{1 \over \epsilon} P_{ad}^{(0)}(z) + {\cal O}(\hat a_s^2)
\end{eqnarray}
where, $P_{ad}^{(0)}(z)$ is the leading order AP splitting function. 
$S_\epsilon = \exp \Big( (\gamma_E-\ln 4 \pi) \frac{\epsilon}{2} \Big)$ 
where $\gamma_E$ is the Euler-Mascheroni constant.
Using $\Gamma_{ad}$, $W^b$ can be written in terms of renormalized ${\cal H}$, given by
 \begin{eqnarray}
{\cal H}_{ac}(x_1,x_2,\mu_F^2) &\equiv& f_a (x_1,\mu_F^2)~ f_c (x_2,\mu_F^2)
\nonumber\\[2ex]
&=&\int_{x_1}^1 {dy_1 \over y_1} \int_{x_2}^1 {dy_2 \over y_2} 
\Gamma_{a a'}(\hat a_s,\mu^2,\mu_F^2,y_1,\epsilon) \hat {\cal H}_{a'c'}\left({x_1\over y_1},{x_2 \over y_2}\right)
\Gamma_{c c'}(\hat a_s,\mu^2,\mu_F^2,y_2,\epsilon) \, .
\nonumber\\
\end{eqnarray}
The LO contribution arises from the Born process $b + \overline b  \rightarrow H$ and the NLO ones are from one loop virtual contributions to born process and from the real emission processes, namely $b + \overline b \rightarrow H+g$, $b(\overline b)+g \rightarrow H+b(\overline b)$. For LO and virtual contributions, $dPS_{1+X}=dPS_1$ and for real emission processes we have two body phase space element $dPS_{1+X}=dPS_2$. In order to define the threshold limit at the partonic level and to express the hadronic cross-section in terms of the partonic one through convolution integrals, we choose to work with the symmetric scaling variables $x_1^0$ and $x_2^0$  instead of $y$ and $\tau$ which are related through
\begin{equation}
 y= \frac{1}{2 } \ln\left( \frac{x_1^0}{x_2^0}\right), \quad \quad \tau = x_1^0 x_2^0 \, .
\end{equation}
In terms of these new variables, the partonic subprocess contributions can be shown to depend on the ratios $z_j = \frac{x_j^0}{x_j}$ which take the role of scaling variables at the partonic level.  The dimensionless partonic differential cross-section denoted by ${\hat \Delta}_{d, ac}^b$ through 
\begin{eqnarray}
{1 \over x_1 x_2} \hat \Delta_{d,ac}^b \left({x_{1}^{0}\over x_1},{x_{2}^{0}\over x_2},\hat a_s,\mu^2,
q^2,\mu_R^2\right) &=&
{(Z^b(\mu_R^2))^2 \over \sigma^{b,(0)}}\int dPS_{1+X} \int dz ~ |{\cal M}_{a c \rightarrow H+X}|^2 ~ 
\nonumber\\[2ex]
&&\times \delta (\tau - z x_1 x_2)
\delta \left(y - \frac{1}{2} \ln \left( \frac{p_2\cdot q}{p_1\cdot q} \right)\right) 
\end{eqnarray}
is UV finite.  Here subscript $d$ stands for differential distribution. The collinear singularities that arise due to the initial state light partons are removed through the AP kernels resulting in the following finite $\Delta_{d,ac}^b$ 
\begin{eqnarray}
\Delta_{d,ac}^b(z_1,z_2,a_s(\mu_R^2),q^2,\mu_F^2,\mu_R^2) & =& 
\int_{z_1}^1 {dy_1\over y_1} \int_{z_2}^1{dy_2\over y_2} \Gamma^{-1}_{aa'}(\hat a_s,\mu^2,\mu_F^2,y_1,\epsilon) 
\nonumber\\[2ex]
&&\hspace{-1cm}\times \hat \Delta_{d,a'c'}^b \left({z_1 \over y_1},{z_2\over y_2},\hat a_s,\mu^2,q^2,\mu_R^2,\epsilon\right)
\Gamma^{-1}_{cc'}(\hat a_s,\mu^2,\mu_F^2,y_2, \epsilon) \, .
\nonumber\\
\end{eqnarray}
Therefore, expressing $W^{b}$ in terms of renormalized ${\cal H}_{ac}$ and finite $\Delta_{d, ac}^b$, we get
\begin{eqnarray}
\label{Wb}
W^b(x_1^0,x_2^0,q^2,\mu_R^2) &=& \sum_{ac=b,\overline b, g} \int_{x_1^0}^1 \frac{d z_1}{z_1} \int_{x_2^0}^1 \frac{d z_2}{z_2} ~ {\cal H}_{ac} \Big( \frac{x_1^0}{z_1}, \frac{x_2^0}{z_2},\mu_F^2 \Big) 
 \nonumber\\[2ex]
&&
 \times \Delta_{d,ac}^b (z_1,z_2,a_s(\mu_R^2),q^2,\mu_F^2,\mu_R^2) \,.
\end{eqnarray}
Since, $W^b$ involves convolutions of various functions, it becomes normal multiplication in the Mellin space of the Mellin moments of renormalized PDFs, AP kernels and bare differential partonic cross-section.  The double Mellin moment of $W^b(x_1^0,x_2^0)$ is defined by 
\begin{eqnarray}
\widetilde{W}^b(N_1,N_2)
&\equiv&\int dx_{1}^{0} \Big(x_{1}^{0}\Big)^{N_1-1}
\int dx_{2}^{0} \Big(x_{2}^{0}\Big)^{N_2-1} W^b(x_{1}^{0},x_{2}^{0})
\nonumber\\[2ex]
&=& \widetilde {\cal H}_{ac}(N_1,N_2) \widetilde \Delta_{d, ac}^b(N_1,N_2)
\end{eqnarray}
where 
\begin{eqnarray}
\widetilde \Delta_{d, ac}^b(N_1,N_2)
&=& \widetilde \Gamma^{-1}_{ae}(N_1)~ \widetilde \Gamma^{-1}_{cf}(N_2)
~\widetilde{ \hat \Delta}_{d, ef}^b(N_1,N_2) \, .
\end{eqnarray}
The threshold limit is defined by $N_i \rightarrow \infty$, which in $z_j$ variables corresponds to $z_j \rightarrow 1$.  In this limit, only diagonal terms in the AP kernel $\widetilde \Gamma^{-1}$ and $\widetilde{\hat \Delta}^b_d$ contribute to the differential cross-section.  Hence, $\ln \tilde{\Delta}^b_d$ is simply a sum of the contributions from 1) diagonal terms of the AP kernels and 2) bare differential partonic cross-section. Due to the born kinematics,  the form factor contribution can be further factored  out from the differential partonic cross sections to all orders in perturbation theory.  Hence, the remaining part of the differential partonic cross-sections contains contributions from only real emission processes, namely those involving only soft gluons.  Taking into account the renormalization constant of the Yukawa coupling ${\hat \lambda}$, we find
\begin{eqnarray}
\ln \widetilde{\Delta}^b_d(N_1,N_2,q^2,\mu_R^2,\mu_F^2) &=& \ln \left(Z^b(\mu_R^2)\right)^2 
-\ln \widetilde{\Gamma}_{bb}(N_1,\mu_F^2) -\ln \widetilde{\Gamma}_{bb}(N_2,\mu_F^2)
\nonumber\\[2ex]
&&+ \ln |{\hat F}^b(q^2)|^2 + \ln \widetilde{S}^b(N_1,N_2,q^2) 
\end{eqnarray}
where $\hat F^b$ and $\widetilde{S}^b(N_1,N_2)$ are bare form factor and real emission contributions of partonic subprocesses, respectively.  The inverse Mellin transform will bring back the expressions in terms of the variables $z_j$ and they will contain besides regular functions, the distributions namely $\delta(1-z_j)$, 
$\D_i$ and $\DD_i$, defined as
\begin{align}
{\cal D}_i=\Bigg[ \frac{\ln^i(1-z_1)}{(1-z_1)} \Bigg]_+ \, , \quad \quad \quad  
\overline {\cal D}_i=\Bigg[ \frac{\ln^i(1-z_2)}{(1-z_2)} \Bigg]_+ \quad \quad \quad i = 0,1,\cdots \, .
\end{align}
The subscript `$+$' denotes the customary `plus-distribution' $f_{+}(z)$ which acts on functions regular in $z \rightarrow 1$ limit as
\begin{equation}
 \int_0^1 dz f_{+}(z) g(z) = \int_0^1 dz f(z) ( g(z) - g(1))
\end{equation}
where, $g(z)$ is any well behaved function in the region $0 \leq z \leq 1$.
In the threshold limit, we drop all the regular terms and keep only these distributions.

In the following, we perform NLO computation in the threshold limit. The overall renormalization constant $(Z^b)^2$ is found to be 
\begin{eqnarray}
(Z^b(\mu_R^2))^2 = 1+ {\hat a_s }S_\epsilon \left({\mu_R^2\over \mu^2}\right)^\epsilon
C_F \left({12 \over \epsilon}\right) + {\cal O}(\hat a_s^2) \, .
\end{eqnarray}
The form factor contribution $|\hat F^b|^2$ at one loop level gives 
\begin{eqnarray}
|\hat F^b(q^2)|^2 = 1+ {\hat a_s }S_\epsilon \left({q^2\over \mu^2}\right)^\epsilon
C_F \left(- \frac{16}{\epsilon}-4 +14 \zeta_2 +{\cal O}(\epsilon) \right)
+{\cal O}(\hat a_s^2)
\end{eqnarray}
The contribution from $\Gamma_{bb}$ in the threshold limit is found to be 
\begin{eqnarray}
&&\int_{z_1}^1 {dy_1\over y_1} \int_{z_2}^1{dy_2\over y_2} \Gamma^{-1}_{bb}(y_1,\mu_F^2) \delta\left(1-{z_1 \over y_1}\right)\delta\left(1-{z_2\over y_2}\right)
\Gamma^{-1}_{bb}(y_2,\mu_F^2)
=\delta(1-z_1) \delta(1-z_2)
\nonumber\\[2ex]
&&\hspace{3cm}-{\hat a_s }S_\epsilon \left({\mu_F^2\over \mu^2}\right)^\epsilon
C_F{1 \over \epsilon}
\Bigg[\left(8 {1 \over (1-z_1)_+} + 6 \delta(1-z_1)\right)\delta(1-z_2)
\nonumber\\[2ex]
&&\hspace{3cm} +\left(8 {1 \over (1-z_2)_+} + 6 \delta(1-z_2)\right)\delta(1-z_1) \Bigg]
+{\cal O}(\hat a_s^2) \, .
\end{eqnarray}
Note that the regular terms in the limit $z_j \rightarrow 1$ in $\Gamma_{bb}$ do not contribute in the threshold limit and hence dropped. 

The inverse Mellin transform of $\tilde S^b(N_1,N_2)$, namely $S^b(z_1,z_2)$ can be obtained directly from the real gluon emission processes in bottom anti-bottom annihilation processes: $b+\overline b \rightarrow
H+g$.  The two body phase space is given by 
\begin{eqnarray}
dPS_{H+g}= {1 \over 8 \pi x_1 x_2} {1 \over \Gamma(1+{\epsilon\over 2})}
\left({m_H^2 \over 4 \pi}\right)^{{\epsilon\over 2}}
{2 z_1 z_2 (1+ z_1 z_2) \over (z_1 + z_2)^{2-\epsilon}} \left((1-z_1^2) (1-z_2^2)
\right)^{{\epsilon \over 2}} \, .
\end{eqnarray}
The phase space in the limit $z_j\rightarrow 1$ becomes
\begin{eqnarray}
dPS_{H+g}|_{z_i \rightarrow 1}= {1 \over 8 \pi x_1 x_2} {1 \over \Gamma(1+{\epsilon\over 2})}
\left({m_H^2 \over 4 \pi}\right)^{{\epsilon\over 2}}
\left((1-z_1^2) (1-z_2^2)
\right)^{{\epsilon \over 2}} \, .
\end{eqnarray}
The spin and color averaged matrix element square in threshold limit is found to be
\begin{eqnarray}
|\overline{M}_{b \overline b \rightarrow H+g}|_{z_j \rightarrow 1}^2
=\sigma_0^b {\hat a_s \over \mu^\epsilon} C_F \Bigg[ {32 \over (1-z_1) (1-z_2)}
+ {\cal O}(\epsilon^3)\Bigg]
\end{eqnarray}
where terms that are regular in $z_j$ as $z_j \rightarrow 1$ have been dropped.
It is then straightforward to obtain the threshold contribution resulting from the real gluon emission process:
\begin{eqnarray}
S^b(z_1,z_2) &=& \delta(1-z_1)\delta(1-z_2)+\hat a_s \left({q^2 \over 4 \pi \mu^2}\right)^{\epsilon \over 2}
{1 \over \Gamma\left(1+{\epsilon\over 2}\right)}
\nonumber\\[2ex]
&& \times 4 C_F \Bigg[ {(1-z_1)^{\epsilon\over 2} (1-z_2)^{\epsilon \over 2} \over (1-z_1) (1-z_2)} + {\cal O}(\epsilon^3)\Bigg] \, .
\end{eqnarray}
Using the identity
\begin{eqnarray}
{(1-z_j)^{a {\epsilon\over 2}} \over (1-z_j)} = {2 \over a \epsilon} \delta(1-z_j)
+ \left({(1-z_j)^{a {\epsilon\over 2}} \over (1-z_j)} \right)_+  \, ,
\label{Defplus}
\end{eqnarray}
it can be shown that $\Delta^b_d(z_1,z_2)$ in the threshold limit contains only the distributions such as $\delta(1-z_j), {\cal D}_i$ and $\overline{\cal D}_i$. Decomposing $\Delta_{d, ac}^b$ into hard and soft parts,
\begin{equation}
\Delta_{d,ac}^b (z_1,z_2,q^2,\mu_F^2,\mu_R^2) = 
     \Delta^{b, {\rm hard}}_{d, ac}(z_1,z_2,q^2,\mu_F^2,\mu_R^2)
   + \delta_{a\overline c} \Delta^{\rm SV}_{d,b}(z_1,z_2,q^2,\mu_F^2,\mu_R^2)\,, 
\end{equation}
and setting $\mu_F=\mu_R=m_H$, we find 
\begin{eqnarray}
\Delta_{d,b}^{{\rm SV},(0)} &=&\delta(1-z_1) \delta(1-z_2) \, ,
\nonumber\\[2ex]
\Delta_{d,b}^{{\rm SV},(1)} &=&
         \delta(1-z_1) \delta(1-z_2) C_F  \Big (
          - 2
          + 6 \zeta_2
          \Big)
       + \D_0 \DD_0   \Big(
            2 C_F
          \Big)
\nonumber\\[2ex]
&&       + \D_1 \delta(1-z_2)    \Big(
           4 C_F
          \Big)
       + \Big\{ z_1 \leftrightarrow z_2 \Big\}
\end{eqnarray}
At the hadronic level, decomposing $W^{b}$ as 
\begin{equation}
\label{WbDecom}
 W^{b}(x_1^0,x_2^0,q^2,\mu_R^2,\mu_F^2) = W^{\rm hard}_{b}(x_1^0,x_2^0,q^2,\mu_R^2,\mu_F^2) +  W^{\rm SV}_{b}(x_1^0,x_2^0,q^2,\mu_R^2,\mu_F^2) \, ,
\end{equation}
similar to $\Delta_{d,b}$ and putting $\Delta_{d,b}^{\rm SV}$ we get, to order $a_s=a_s(m_H^2)$
\begin{eqnarray}
\label{WbSV}
W^{{\rm SV}}_{b}(x_1^0,x_2^0,q^2,m_H^2)& =& {\cal H}_{b \overline b}(x_1^0,x_2^0)
 + a_s 4 C_F \Bigg[{\cal H}_{b \overline b}(x_1^0,x_2^0) \Bigg(
-1 + \zeta_2 + li_2(x_1^0) +li_2(x_2^0)
\nonumber\\[2ex]
&&+{1 \over 2} \ln^2\left((1-x_1^0)(1-x_2^0)\right)
+\ln\left((1-x_1^0)\over x_2^0\right)
\ln\left((1-x_2^0)\over x_1^0\right)\Bigg)
\nonumber\\[2ex]
&&+ \int dx_1 {\cal H}_{b \overline b,1} (x_1,x_2^0)
{1 \over x_1-x_1^0} \ln\left((1-x_2^0)(x_1-x_1^0) \over x_1 x_2^0\right)
\nonumber\\[2ex]
&&+ \int dx_2 {\cal H}_{b \overline b,1} (x_1^0,x_2)
{1 \over x_2-x_2^0} \ln\left((1-x_1^0)(x_2-x_2^0) \over x_1^0 x_2\right)
\nonumber\\[2ex]
&&+ \int dx_1 \int dx_2 {\cal H}_{b \overline b,12} (x_1,x_2)
{1 \over (x_1 - x_1^0) (x_2-x_2^0)}\Bigg] \, .
\end{eqnarray}
where all the parton densities are defined at $\mu_F=m_H$.  In general,
\begin{align}
 {\cal H}_{b \overline{b},12}(x_1,x_2,\mu_F^2) &\equiv {\cal H}_{b \overline{b}}(x_1,x_2,\mu_F^2) - {\cal H}_{b \overline{b}}(x_1^0,x_2,\mu_F^2) - {\cal H}_{b \overline{b}}(x_1,x_2^0,\mu_F^2) + {\cal H}_{b \overline{b}}(x_1^0,x_2^0,\mu_F^2),
\nonumber\\
 {\cal H}_{b \overline{b},1} (x_1,x_2,\mu_F^2) &\equiv {\cal H}_{b \overline{b}}(x_1,x_2,\mu_F^2) - {\cal H}_{b \overline{b}}(x_1^0,x_2,\mu_F^2),
\nonumber\\
 {\cal H}_{b \overline{b},2} (x_1,x_2,\mu_F^2) &\equiv {\cal H}_{b \overline{b}}(x_1,x_2,\mu_F^2) - {\cal H}_{b \overline{b}}(x_1,x_2^0,\mu_F^2)
\end{align}
The Spence function ($li_2 (x)$) is defined as
\begin{eqnarray}
li_2(x) \equiv -\int_0^x {d z \over z} \ln(1-z) \, .
\end{eqnarray} 

The exact result computed at NLO level confirms our expectations, see for example \cite{Choudhury:2005eu} where the rapidity distribution of di-leptons in the Drell-Yan production for a physics beyond the SM (BSM) involving a generic Yukawa type interaction was obtained to NLO level.  After the suitable replacement of the BSM coupling in \cite{Choudhury:2005eu}, we obtain 
\begin{equation}
\frac{d\sigma^b}{dy} (\tau, y, Q^2) =  \sigma^{b,(0)}(\mu_F) \Bigg[ W_{b\bar{b}} (x_1^0,x_2^0,\mu_F^2)
           + W_{bg} (x_1^0,x_2^0,\mu_F^2) + W_{gb} (x_1^0,x_2^0,\mu_F^2) \Bigg]
\end{equation}
where $W$'s can be expanded in the strong coupling constant $a_s(\mu_F^2)$ as 
\begin{equation}
 W_{ac} (x_1^0,x_2^0,\mu_F^2) = W_{ac}^{(0)} (x_1^0,x_2^0,\mu_F^2) + a_s(\mu_F^2) W_{ac}^{(1)} (x_1^0,x_2^0,\mu_F^2) + {\cal O} (a_s^2) 
\end{equation}
and the corresponding coefficients are given by 
\begin{align}
 W_{b\bar{b}}^{(0)} (x_1^0,x_2^0,\mu_F^2) &=  {\cal H}_{b\bar{b}} (x_1^0,x_2^0,\mu_F^2)
\nonumber\\
W_{b\bar{b}}^{(1)} (x_1^0,x_2^0,\mu_F^2) &=  {2 \, \, C_F} \; \Bigg\{ \varphi^{b \bar b}_0 + 
    \int dx_1 \,  \varphi^{b \bar b}_1 + \int dx_1 dx_2 \, \varphi^{b \bar b}_2  \Bigg\} + \Big(1 \leftrightarrow 2 \Big)
\nonumber\\
W_{gb}^{(1)} (x_1^0,x_2^0,\mu_F^2) &=  {2 \, T_f} \int \frac{dx_1}{x_1^3} \Bigg[ \varphi^{g \bar b}_1 + \int dx_2  \left\{ \varphi^{g \bar b}_2 -  
 \frac{\varphi^{g \bar b}_3 \; {\cal H}_{gb}(x_1,x_2,\mu_F^2) }{ x_2^2 \, (x_2+x_2^0) \, (x_1 x_2^0 + x_2 x_1^0)^3} \right\} \Bigg]
\nonumber\\
W_{bg}^{(1)} (x_1^0,x_2^0,\mu_F^2) &= W_{gb}^{(1)} (x_1^0,x_2^0,\mu_F^2) |_{(1 \leftrightarrow 2)}
\end{align}
with
\begin{align}
 \varphi^{b \bar b}_0  &= \frac{1}{2} {\cal H}_{b\bar{b}} (x_1^0,x_2^0,\mu_F^2)  
     \Bigg( -2 + \kappa_{12}^2 + 6 \, \zeta_2 +  2\, \kappa_{12} 
     \ln \, \frac{q^2}{\mu^2_F} \Bigg)
\nonumber\\
 \varphi^{b \bar b}_1 &= \frac{2 \kappa_{b_1}}{x_1 - x_1^0} {\cal H}_{b\bar{b},1} (x_1,x_2^0,\mu_F^2)
+ {\cal H}_{b\bar{b}} (x_1,x_2^0,\mu_F^2) \Bigg( \frac{1-\kappa_{a_1}}{x_1} + \frac{2 \kappa_{c_1}}{x_1-x_1^0} - \frac{1+\kappa_{a_1}}{x_1^2} x_1^0\Bigg)
\nonumber\\
 \varphi^{b \bar b}_2 &= \frac{{\cal H}_{b\bar{b},12} (x_1,x_2,\mu_F^2)}{(x_1-x_1^0) (x_2-x_2^0)}
   - \frac{x_2+x_2^0}{(x_1-x_1^0)  x_2^2}  {\cal H}_{b\bar{b},1} (x_1,x_2,\mu_F^2)
\nonumber\\
 &+ \frac{{\cal H}_{b\bar{b}} (x_1,x_2,\mu_F^2)}{2 x_1^2 x_2^2 } \Bigg( (x_1+x_1^0) \, (x_2+x_2^0)
 + \frac{x_1^2 x_2^2 + x_1^{0^2} \, x_2^{0^2}}{(x_1+x_1^0) \, (x_2+x_2^0)} \Bigg)
\nonumber\\
 \varphi^{g \bar b}_1  &= {\cal H}_{gb} (x_1,x_2^0,\mu_F^2) \Bigg( 2 x_1^0 (x_1-x_1^0) + \kappa_{a_1} \Big(x_1^{0^2} + (x_1-x_1^0)^2\Big)\Bigg)
\nonumber\\
 \varphi^{g \bar b}_2  &= \frac{{\cal H}_{gb,2} (x_1,x_2,\mu_F^2)}{x_2-x_2^0 } \Big( x_1^{0^2} + (x_1-x_1^0)^2\Big)
\nonumber\\
 \varphi^{g \bar b}_3  &= - x_1^5 x_2^2 x_2^{0^3} + x_1^4 x_1^0 x_2^2 x_2^{0^2} (3 x_2 + 4 x_2^0)
+ x_1^3 x_1^{0^2} x_2 x_2^0 (3 x_2^3 + 2 x_2^{0^3}) + 2 x_1^{0^5} x_2^2
(x_2^3 + 2 x_2^2 x_2^0 
\nonumber\\
&
+ 2 x_2 x_2^{0^2} + 2x_2^{0^3}) 
+ 2 x_1 x_1^{0^4} x_2
(-x_2^4 + x_2^3 x_2^0 + 4 x_2^2 x_2^{0^2} + 2 x_2 x_2^{0^3} + 2 x_2^{0^4})
\nonumber\\
&
+ x_1^2 x_1^{0^3} (x_2^5 - 4 x_2^4 x_2^0 -4 x_2^3 x_2^{0^2} + 
2 x_2^2 x_2^{0^3} + 2 x_2 x_2^{0^4} + 2 x_2^{0^5}) 
\end{align}
and
\begin{align}
 &{\cal \kappa}_{a_1} = \ln \, \frac{2 \, q^2 \, (1-x_2^0) \, (x_1-x_1^0)}{\mu_F^2 \, (x_1+x_1^0) \, x_2^0} \,,
% \nonumber\\
% % 
%  \qquad
 &{\cal \kappa}_{b_1} = \ln \, \frac{ q^2 \, (1-x_2^0) \, (x_1-x_1^0)}{\mu_F^2 \, x_1^0 \, x_2^0} \,
\nonumber\\
 &{\cal \kappa}_{c_1} = \ln \, \frac{2 \, x_1^0}{x_1 + x_1^0} 
% \nonumber\\
% 
 &{\cal \kappa}_{12} = \ln \, \frac{ (1-x_1^0) \, (1-x_2^0)}{x_1^0 \, x_2^0} \qquad \, .
\end{align}
In the threshold limit, after setting $\mu_F=m_H$, we find that the above result reduces to given in Eq.~\ref{WbSV}. 

%************************************
% Threshold corrections beyond NLO
%************************************

\subsection{Threshold Corrections Beyond NLO}
\label{ThresBNLO}

Following the factorization approach that we used in the previous section to obtain the threshold correction to NLO rapidity distribution, we now set up a framework to compute threshold correections to rapidity distribution  to all orders in strong coupling constant.   Our approach is based on the fact that the rapidity distribution in the threshold limit can be systematically factorized into 1) the exact form factor, 2) overall UV renormalization constant, 3) soft gluon contributions from real emission partonic subprocesses and 4) the diagonal collinear subtraction terms  involving  only $\delta(1-z)$ and ${\cal D}_{0}(z)$ terms of AP splitting functions. We call such a combination soft-virtual (SV) part of the rapidity distribution and the remaining part as hard. Hence, we propose that 
\begin{equation}
\Delta^{\rm SV}_{d,b}(z_1,z_2,q^2,\mu_R^2,\mu_F^2) = {\cal C} \exp \Bigg({\Psi^b_d(q^2,\mu_R^2,\mu_F^2,z_1,z_2,\ep)}\Bigg)\Bigg|_{\ep=0}\, .
\label{master}
\end{equation}
The symbol `${\cal C}$' means convolution with the following definition 
\begin{align}
{\cal C}e^{\displaystyle f(z_1,z_2) } &= \delta(1-z_1)\delta(1-z_2)  + \frac{1}{1!} f(z_1,z_2) + \frac{1}{2!} f(z_1,z_2) \otimes f(z_1,z_2) 
\nonumber\\
&
+ \frac{1}{3!} f(z_1,z_2) \otimes f(z_1,z_2) \otimes f(z_1,z_2) + \cdots \,,
\label{convexp}
\end{align}
where, $\otimes$ indicates double Mellin convolution with respect to the variables $z_1$ and $z_2$ and the function $f(z_1,z_2)$ is a distribution of the kind $\delta(1-z_j)$ and/or ${\cal D}_i (z_j)$. The finite distribution $\Psi^b_d$ in dimensional regularization contains $Hb\bar{b}$ unrenormalized form factor $\hat{F}^b(\hat a_s,q^2=-Q^2,\mu^2,\ep)$, UV overall operator renormalization constant $Z^b(\hat a_s,\mu_R^2,\mu^2,\ep)$, soft distribution functions $\Phi^{b}_d(\hat a_s,q^2,\mu^2,z_1,z_2,\ep)$ and the mass factorization kernels $\Gamma_{bb}(\hat a_s,\mu^2,\mu_F^2,z_j,\ep)$:
\begin{align}
 \Psi^b_d &=
% \Psi^b_d(q^2,\mu_R^2,\mu_F^2,z_1,z_2,\ep) &=
   \Bigg( \ln \Big( Z^b(\hat{a}_s,\mu_R^2,\mu^2,\ep) \Big)^2 + \ln \big|\hat F^b(\hat a_s,Q^2,\mu^2,\ep)\big|^2 \Bigg) \delta(1-z_1) \delta(1-z_2)
\nonumber\\
& 
  + 2~ \Phi^{b}_d(\hat a_s,q^2,\mu^2,z_1,z_2,\ep) - {\cal C}\ln \Gamma_{bb}(\hat a_s,\mu^2,\mu_F^2,z_1,\ep)~ \delta(1-z_2)
\nonumber\\[2ex]
& 
  - {\cal C}\ln \Gamma_{bb}(\hat a_s,\mu^2,\mu_F^2,z_2,\ep)~ \delta(1-z_1) \,.
\label{DYH}
\end{align}

%%%% Renormalization

We have expressed all the quantities in the above equation in terms of unrenormalized strong coupling constant $\hat a_s$ related to the standard $\hat{\alpha}_{s}$ through ${\hat a}_{s} = \hat{\alpha}_{s}/{4\,\pi}$ and the dimensional regularization scale $\mu$.  The UV renormalization of ${\hat a}_s$ is done at the renormalization scale $\mu_R$ through $Z (\mu_R^2)$ giving the renormalized $a_s (\mu_R^2)$, that is   
\begin{eqnarray}
\hat a_s  = \Big( \frac{\mu}{\mu_R} \Big)^\epsilon 
Z (\mu_R^2) S_\epsilon^{-1} a_s(\mu_R^2).  
\end{eqnarray}
 The renormalization group equation (RGE) for $a_s(\mu_R^2)$ 
\begin{eqnarray}
\mu_R^2 \frac{d a_s(\mu_R^2)}{d \mu_R^2} &=&
      \frac{\epsilon ~ a_s(\mu_R^2)}{2} + \beta (a_s(\mu_R^2)) 
\label{rge1}
\end{eqnarray}
with
\begin{eqnarray}
\beta (a_s(\mu_R^2))= 
  a_s(\mu_R^2)\, \mu_R^2  \frac{d \ln Z (\mu_R^2)}{d \mu_R^2}
= - \sum_{i=0}^{\infty} a_s^{i+2} (\mu_R^2) \beta_i \,,
\end{eqnarray}
determines the structure of the $Z(\mu_R^2)$, up to ${\cal O}(a_s^3)$, we find
\begin{equation}
Z(\mu_R^2)= 1+ a_s(\mu_R^2) \frac{2}{\ep} \beta_0 
           + a_s^2(\mu_R^2) \Bigg(\frac{4}{\ep^2 } \beta_0^2  +
                  \frac{1}{\ep} \beta_1 \Bigg)
           + a_s^3(\mu_R^2) \Bigg( \frac{8}{ \ep^3} \beta_0^3
                   +\frac{14}{3 \ep^2} \beta_0 \beta_1 
                   + \frac{2}{3 \ep}  \beta_2 \Bigg) .~~~~
\end{equation}
The first three coefficients of the QCD $\beta$ function, $\beta_0$, $\beta_1$ and $\beta_2$ are given by \cite{Tarasov:1980au} 
\begin{align}
\beta_0&={11 \over 3 } C_A - {4 \over 3 } T_F n_f \, ,
\nonumber \\[0.5ex]
\beta_1&={34 \over 3 } C_A^2-4 T_F n_f C_F -{20 \over 3} T_F n_f C_A \, ,
\nonumber \\[0.5ex]
\beta_2&={2857 \over 54} C_A^3 
          -{1415 \over 27} C_A^2 T_F n_f
          +{158 \over 27} C_A T_F^2 n_f^2
\nonumber\\[0.5ex]
&          +{44 \over 9} C_F T_F^2 n_f^2
          -{205 \over 9} C_F C_A T_F n_f
          +2 C_F^2 T_F n_f  
\end{align}
with the $SU(N)$ color factors 
\begin{equation}
C_A=N,\quad \quad \quad C_F= \frac{N^2-1}{2 N} , \quad \quad \quad
T_F= \frac{1}{2}
\end{equation}
and $n_f$ is the number of active flavours.

The overall operator renormalization constant $Z^b$ renormalizes the bare Yukawa coupling $\hat{\lambda}$ resulting $\lambda (\mu_R^2)$ through the relation
\begin{equation}
\label{RGElambda}
 \hat{\lambda} = \Big( \frac{\mu}{\mu_R} \Big)^{\epsilon\over 2}
Z^b(\mu_R^2) S_{\epsilon}^{-1}
\lambda (\mu_R^2)\,.
\end{equation}
In $\overline{\rm{MS}}$ scheme, $Z^b(\mu_R^2)$ is identical to quark mass renormalization constant. The RGE for $\lambda(\mu_R^2)$ takes the form 
\begin{equation}
 \mu_R^2 \frac{d}{d\mu_R^2} \ln Z^b (\hat{a}_s, \mu_R^2, \mu^2, \epsilon) = \sum_{i=1}^{\infty}  a_s^i (\mu_R^2) \gamma^b_{i-1} \, ,
\end{equation}
with the anomalous dimensions $\gamma^b_i$ given by \cite{Tarasov:1982gk, vanRitbergen:1997va, Czakon:2004bu}
\begin{align}
\gamma^b_0&= 3 C_F \, ,
\nonumber \\[0.5ex]
\gamma^b_1&= \frac{3}{2} C_F^2
           + \frac{97}{6} C_F C_A
           - \frac{10}{3} C_F T_F n_f \, ,
\nonumber \\[0.5ex]
\gamma^b_2&= \frac{129}{2} C_F^3 
           - \frac{129}{4} C_F^2 C_A
           + \frac{11413}{108} C_F C_A^2
           +\Big(-46+48 \zeta_3\Big) C_F^2 T_F n_f
\nonumber \\[0.5ex]
&           +\left(-\frac{556}{27} -48 \zeta_3\right) C_F C_A T_F n_f
           - \frac{140}{27} C_F T_F^2 n_f^2 \, .
\end{align}
Upon solving the above RGE in $4+\epsilon$ space-time dimensions, we obtain
\begin{align}
 \ln Z^b(\mu_R^2) &=
         a_s(\mu_R^2) \frac{1}{\epsilon}   \Bigg(
            2 {\gamma^{b}_{0}}
          \Bigg)
       + a_s^2(\mu_R^2) \Bigg[
         \frac{1}{\epsilon^2}   \Bigg(
             2 \bt_0 {\gamma^{b}_{0}}
          \Bigg)
       + \frac{1}{\epsilon}   \Bigg(
             {\gamma^{b}_{1}}
          \Bigg)\
       \Bigg]      
\nonumber\\
&
       + a_s^3(\mu_R^2) \Bigg[
         \frac{1}{\epsilon^3}   \Bigg(
            \frac{8}{3} \bt_0^2 {\gamma^{b}_{0}}
          \Bigg)
       + \frac{1}{\epsilon^2}  \Bigg(
            \frac{4}{3} \bt_1 {\gamma^{b}_{0}}
          + \frac{4}{3} \bt_0 {\gamma^{b}_{1}}
          \Bigg)
       + \frac{1}{\epsilon}   \Bigg(
            \frac{2}{3} {\gamma^{b}_{2}}
          \Bigg)
       \Bigg]
\label{logZb}
\end{align}
up to ${\cal O}(a_s^3)$. 

The bare form factor $\hat{F}^b (\hat{a}_s, Q^2, \mu^2, \epsilon)$ satisfies the following differential equation which follows from the gauge as well as renormalization group invariances \cite{Sudakov:1954sw, Mueller:1979ih, Collins:1980ih, Sen:1981sd} 
\begin{equation}
 Q^2 \frac{d}{dQ^2} \ln \hat{F}^b  = \frac{1}{2} \Big[ K^b (\hat{a}_s, \frac{\mu_R^2}{\mu^2}, \epsilon ) + G^b (\hat{a}_s, \frac{Q^2}{\mu_R^2}, \frac{\mu_R^2}{\mu^2}, \epsilon ) \Big]
 \label{rgeF}
\end{equation}
where, all the poles in $\epsilon$ are encapsulated within $K^b$  and $G^b$ contains the terms finite in $\epsilon$. Renormalization group invariance of  $\hat{F}^b (\hat{a}_s, Q^2, \mu^2, \epsilon)$ leads
\begin{equation}
\mu_R^2 \frac{d}{d\mu_R^2} K^b = - \mu_R^2 \frac{d}{d\mu_R^2} G^b 
= - \sum_{i=1}^{\infty}  a_s^i (\mu_R^2) A^q_i \,,
\label{rgeKG}
\end{equation}
where, $A^q_i$'s are the cusp anomalous dimensions, found to be \cite{Moch:2004pa, Vogt:2004mw, Catani:1989ne, Catani:1990rp, Vogt:2000ci}
 \begin{align}
  A^q_1 &= 4 C_F \,, 
 \nonumber \\
  A^q_2 &= 8 C_F C_A \Bigg\{ \frac{67}{18} - \zeta_2 \Bigg\} + 8 C_F n_f \Bigg\{ -\frac{5}{9} \Bigg\} \,,
 \nonumber \\
  A^q_3 &= 16 C_F C_A^2 \Bigg\{ \frac{245}{24} - \frac{67}{9} \zeta_2  + \frac{11}{6} \zeta_3
                              + \frac{11}{5} \zeta_2^2 \Bigg\}
             + 16 C_F^2 n_f \Bigg\{ - \frac{55}{24} + 2 \zeta_3 \Bigg\}
 \nonumber\\
 &
             + 16 C_F C_A n_f \Bigg\{ - \frac{209}{108} + \frac{10}{9} \zeta_2 - \frac{7}{3} \zeta_3 \Bigg\}
             + 16 C_F n_f^2 \Bigg\{ - \frac{1}{27} \Bigg\} \,.
 \end{align}
Being flavor independent, $A_{i}^{b}$'s are same as $A_{i}^{q}$. Solving the RGE \ref{rgeKG} satisfied by $K^{b}$  we get
\begin{eqnarray}
K^b(\hat a_s,\mu^2,\mu_R^2,\epsilon) = \sum_{i=1}^\infty \hat a_s^i
\left({\mu_R^2 \over \mu^2}\right)^{i {\epsilon\over 2}} S_\epsilon^i
K^{b,(i)}(\epsilon)
\label{Kbexp}
\end{eqnarray}
with
\begin{eqnarray}
K^{b,(1)}(\epsilon)&=& {1 \over \epsilon} \Bigg\{ -2 A_1^{b}\Bigg\},\quad 
K^{b,(2)}(\epsilon)= {1 \over \epsilon^2} \Bigg\{ 2 \beta_0 A_1^{b} \Bigg\}
+{1 \over \epsilon} \Bigg\{ - A_2^{b} \Bigg\} \,,
\nonumber\\[2ex]
K^{b,(3)}(\epsilon)&=& { 1 \over \epsilon^3} \Bigg\{ -{8 \over 3} \beta_0^2 A_1^{b} \Bigg\}
+{1 \over \epsilon^2} \Bigg\{ {2 \over 3} \beta_1 A_1^{b} 
+{8 \over 3} \beta_0 A_2^{b}\Bigg\} + {1 \over \epsilon} \Bigg\{ -{2 \over 3} A_3^{b} \Bigg\} \, .
\end{eqnarray}
Similarly upon solving the RGE \ref{rgeKG} for $G^b$, we obtain
\begin{align}
\label{Gbexp}
 G^b (\hat{a}_s, \frac{Q^2}{\mu_R^2}, \frac{\mu_R^2}{\mu^2}, \epsilon ) 
&   = G^b (a_s (\mu_R^2), \frac{Q^2}{\mu_R^2}, \epsilon ) 
\nonumber\\
&   = G^b (a_s (Q^2), 1, \epsilon ) + \int_{Q^2/\mu_R^2}^{1} \frac{d \lambda^2}{\lambda^2} A^{b} (a_s(\lambda^2 \mu_R^2))
\nonumber\\
&   = G^b (a_s (Q^2), 1, \epsilon ) + \sum_{i=1}^{\infty} S_{\epsilon}^i \hat{a}_s^i \Big( \frac{\mu_R^2}{\mu^2} \Big)^{i \frac{\epsilon}{2}} 
                                                 \Big[ \Big( \frac{Q^2}{\mu_R^2} \Big)^{i \frac{\epsilon}{2}} - 1 \Big] K^{b,(i)} (\epsilon)\, .
\end{align}
Expanding the finite function $G^b (a_s (Q^2), 1, \epsilon )$ in powers of $a_s(Q^2)$ as
\begin{equation}
 G^b (a_s (Q^2), 1, \epsilon ) = \sum_{i=1}^{\infty} a_s^i (Q^2) G_i^b (\epsilon) \, ,
\end{equation}
one finds that $G^b_i$ can be expressed in terms of collinear $B^q_i$ and soft $f^q_i$ anomalous dimensions through the relation \cite{Ravindran:2004mb, Becher:2009cu, Gardi:2009qi}
\begin{eqnarray}
 G^b_i (\epsilon) = 2 (B^q_i - \gamma^b_i) + f^q_i + C^b_i + \sum_{k=1}^{\infty} \epsilon^k g_i^{b,k} \,.
\label{Gbexp}
\end{eqnarray}
Note that the single pole term of the form factor depends on three different anomalous dimensions, namely the collinear anomalous dimension $B^q_i$,
anomalous dimension of the coupling constant $\gamma^b_i$ and the soft anomalous dimension $f_i^q$.  $B^q_i$ can be obtained from the $\delta(1-z)$ part of the diagonal splitting function known up to three loop level \cite{Moch:2004pa, Vogt:2004mw} which are   
\begin{align}
 B^q_1 &= 3 C_F \,,
 \nonumber \\
 B^q_2 &=  \frac{1}{2} \Bigg( C_F^2 \Bigg\{ 3 - 24 \zeta_2 + 48 \zeta_3 \Bigg\}
              + C_A C_F \Bigg\{ \frac{17}{3} + \frac{88}{3} \zeta_2 - 24 \zeta_3 \Bigg\}
                + n_f T_F C_F \Bigg\{ - \frac{4}{3} - \frac{32}{3} \zeta_2 \Bigg\} \Bigg) \,,
 \nonumber \\
 B^q_3 &= - 16 {C_A}^2 {C_F} \Bigg\{ \frac{1}{8} {\zeta_2}^2 - \frac{281}{27} {\zeta_2}
          + \frac{97}{9} {\zeta_3} - \frac{5}{2} {\zeta_5} + \frac{1657}{576} \Bigg\}
          + 16 {C_A} {C_F}^2 \Bigg\{ -\frac{247}{60} {\zeta_2}^2 + {\zeta_2} {\zeta_3} 
\nonumber\\
&
          - \frac{205}{24} {\zeta_2} +\frac{211}{12} {\zeta_3}
          + \frac{15}{2} {\zeta_5} + \frac{151}{64} \Bigg\}
+16 {C_A} {C_F} {n_f} \Bigg\{ \frac{1}{20} {\zeta_2}^2 - \frac{167}{54} {\zeta_2}
          + \frac{25}{18} {\zeta_3} + \frac{5}{4} \Bigg\}
\nonumber\\
&
+16 {C_F}^3 \Bigg\{ \frac{18}{5} {\zeta_2}^2 - 2 {\zeta_2} {\zeta_3}
          +\frac{9}{8} {\zeta_2} + \frac{17}{4} {\zeta_3} - 15 {\zeta_5} + \frac{29}{32} \Bigg\}
\nonumber\\
&
%           %           
-16 {C_F}^2 {n_f} \Bigg\{ - \frac{29}{30} {\zeta_2}^2 - \frac{5}{12} {\zeta_2}
          +\frac{17}{6} {\zeta_3} + \frac{23}{16} \Bigg\}
-16 {C_F} {n_f}^2 \Bigg\{ -\frac{5}{27} {\zeta_2} + \frac{1}{9} {\zeta_3} + \frac{17}{144} \Bigg\} \, .
\end{align}
The $f_i^q$ for $i=1,2$ can be found in \cite{Ravindran:2004mb} and in \cite{Moch:2004pa} for $i=3$. We list them below:
\begin{align}
 f_1^q &= 0 \,,
\nonumber \\
 f_2^q &= C_A C_F \Bigg\{ -\frac{22}{3} {\zeta_2} - 28 {\zeta_3} + \frac{808}{27} \Bigg\}
        + C_F n_f T_F \Bigg\{ \frac{8}{3} {\zeta_2} - \frac{224}{27} \Bigg\} \,,
\nonumber \\
 f_3^q &= {C_A}^2 C_F \Bigg\{ \frac{352}{5} {\zeta_2}^2 + \frac{176}{3} {\zeta_2} {\zeta_3}
        - \frac{12650}{81} {\zeta_2} - \frac{1316}{3} {\zeta_3} + 192 {\zeta_5}
        + \frac{136781}{729}\Bigg\}
\nonumber \\
&
        + {C_A} {C_F} {n_f} \Bigg\{ - \frac{96}{5} {\zeta_2}^2 
        + \frac{2828}{81} {\zeta_2}
        + \frac{728}{27} {\zeta_3} - \frac{11842}{729} \Bigg\} 
\nonumber \\
&
        + {C_F}^2 {n_f} \Bigg\{ \frac{32}{5} {\zeta_2}^2 + 4 {\zeta_2} 
        + \frac{304}{9} {\zeta_3} - \frac{1711}{27} \Bigg\}
        + {C_F} {n_f}^2 \Bigg\{ - \frac{40}{27} {\zeta_2} + \frac{112}{27} {\zeta_3}
        - \frac{2080}{729} \Bigg\} \, .
\end{align}
Since $B^q_i$ and $f_i^q$ are flavour independent, we have used $B^b_i \equiv B^q_i$ and $f_i^b \equiv f_i^q$ in $G^b_i$.
The constants $C^b_i$ are controlled by the beta function of the strong coupling constant through renormalization group invariance of the bare form factor:
\begin{equation}
C^b_1 = 0,\quad \quad  C^b_2 = - 2 \beta_0 g_1^{b,1}, \quad \quad
% \nonumber\\[2ex]
C^b_3 = - 2 \beta_1 g_1^{b,1} - 2 \beta_0 ( g_2^{b,1} + 2 \beta_0  g_1^{b,2}). 
\end{equation}
The coefficients $g_i^{b, k}$ can be extracted from the finite part of the form factor. Up to two loop level, we use \cite{Harlander:2003ai, Ravindran:2005vv, Ravindran:2006cg} and at three loop level the recent computation by Gehrmann and Kara \cite{Gehrmann:2014vha} enable us to compute the relevant $g_3^{b,1}$ in \cite{Ahmed:2014cha} where $g_3^{b,1}$ was already used to obtain threshold correction to inclusive Higgs production in bottom anti-bottom annihilation process:
\begin{align}
 g_1^{b,1} &= C_F \Bigg\{ - 2 + \zeta_2 \Bigg\} ,\, \quad 
 g_1^{b,2} = C_F \Bigg\{ 2 - \frac{7}{3} \zeta_3 \Bigg\} , \,\quad 
 g_1^{b,3} = C_F \Bigg\{ - 2 + \frac{1}{4} \zeta_2 + \frac{47}{80} \zeta_2^2 \Bigg\} \,,\nonumber \\
g_2^{b,1} &= C_F n_f \Bigg\{ \frac{616}{81} + \frac{10}{9} \zeta_2 - \frac{8}{3} \zeta_3 \Bigg\}
          + C_F C_A \Bigg\{ - \frac{2122}{81} - \frac{103}{9} \zeta_2 + \frac{88}{5} {\zeta_2}^2 + \frac{152}{3} \zeta_3 \Bigg\}
\nonumber \\
         & + C_F^2 \Bigg\{ 8 + 32 \zeta_2 - \frac{88}{5} {\zeta_2}^2 - 60 \zeta_3 \Bigg\}  \,,
\nonumber \\
g_2^{b,2} &= C_F n_f \Bigg\{ \frac{7}{12} {\zeta_2}^2 - \frac{55}{27} \zeta_2 + \frac{130}{27} \zeta_3 - \frac{3100}{243} \Bigg\}
           + C_A C_F  \Bigg\{ - \frac{365}{24} {\zeta_2}^2 + \frac{89}{3} \zeta_2 \zeta_3 + \frac{1079}{54} \zeta_2 
\nonumber \\
         & - \frac{2923}{27} \zeta_3 - 51 \zeta_5 + \frac{9142}{243} \Bigg\}
           + C_F^2 \Bigg\{ \frac{ 96}{5} {\zeta_2}^2 - 28 \zeta_2 \zeta_3 
           - 44 \zeta_2 + 116 \zeta_3 + 12 \zeta_5 - 24 \Bigg\} \, ,
\nonumber \\%
g_3^{b,1} &= C_A^2 C_F   \Bigg\{ - \frac{6152}{63} {\zeta_2}^3 + \frac{2738}{9} {\zeta_2}^2
 + \frac{976}{9} \zeta_2 \zeta_3 - \frac{342263}{486} \zeta_2
 - \frac{1136}{3} {\zeta_3}^2 + \frac{19582}{9} \zeta_3 
\nonumber \\
&
 + \frac{1228}{3} \zeta_5 
 + \frac{4095263}{8748} \Bigg\}
+ C_A C_F^2  \Bigg\{ - \frac{15448}{105} {\zeta_2}^3 - \frac{3634}{45} {\zeta_2}^2
 - \frac{2584}{3} \zeta_2 \zeta_3 + \frac{13357}{9} \zeta_2 
\nonumber \\
&
 + 296 \zeta_3^2
 - \frac{11570}{9} \zeta_3 - \frac{1940}{3} \zeta_5 - \frac{613}{3} \Bigg\}
+ C_A C_F n_f  \Bigg\{ - \frac{1064}{45} {\zeta_2}^2 + \frac{392}{9} \zeta_2 \zeta_3 
 + \frac{44551}{243} \zeta_2 
\nonumber \\
&
 - \frac{41552}{81} \zeta_3 
 - 72 \zeta_5 - \frac{6119}{4374} \Bigg\}
+ C_F^2 n_f  \Bigg\{ \frac{772}{45} {\zeta_2}^2 - \frac{152}{3} \zeta_2 \zeta_3
  - \frac{3173}{18} \zeta_2 + \frac{15956}{27} \zeta_3 -\frac{368}{3} \zeta_5
\nonumber \\
&  
  + \frac{32899}{324}\Bigg\}
+ C_F n_f^2  \Bigg\{ - \frac{40}{9} {\zeta_2}^2 - \frac{892}{81} \zeta_2 
  + \frac{320}{81} \zeta_3 - \frac{27352}{2187} \Bigg\}
+ C_F^3 \Bigg\{ \frac{21584}{105} {\zeta_2}^3 - \frac{1644}{5} {\zeta_2}^2
\nonumber \\
&
  + 624 \zeta_2 \zeta_3 
  - 275 \zeta_2 + 48 \zeta_3^2 
  - 2142 \zeta_3 + 1272 \zeta_5 + 603 \Bigg\} \,.
\end{align}
 Using the expressions for $K^b$ and $G^b$ given in Eq.~\ref{Kbexp} and Eq.~\ref{Gbexp}, respectively, we obtain the renormalized form factor up to order ${\cal O}(a_s^3)$ as
\begin{align}
\ln |{\hat F}^{b}|^2 (Q^2, \epsilon) &=
         a_s(q^2) \Bigg[
         \frac{1}{\epsilon^2}   \Big(
          - 4 {A^{q}_{1}}
          \Big)
       + \frac{1}{\epsilon}   \Big(
            2 {f^{q}_{1}}
          + 4 {B^{q}_{1}}
          - 4 {\gamma^{b}_{0}}
          \Big)
       + \Big(
            2 {g^{b,1}_{1}}
          + 3 \zeta_2 {A^{q}_{1}}
          \Big)
       \Bigg]
\nonumber\\
& 
       + a_s^2 (q^2) \Bigg[
         \frac{1}{\epsilon^3}   \Big(
          - 6 \bt_0 {A^{q}_{1}}
          \Big)
       + \frac{1}{\epsilon^2}   \Big(
          - {A^{q}_{2}}
          + 2 \bt_0\Big(
            {f^{q}_{1}}
          + 2  {B^{q}_{1}}
          - 2 {\gamma^{b}_{0}} \Big)
          \Big)
       + \frac{1}{\epsilon}   \Big(
            {f^{q}_{2}}
\nonumber\\
& 
          + 2 {B^{q}_{2}}
          - 2 {\gamma^{b}_{1}}
          \Big)
       + \Big(
            {g^{b,1}_{2}}
          + 2 \bt_0 {g^{b,2}_{1}}
          + 3 \zeta_2 {A^{q}_{2}}
          + 3 \zeta_2 \bt_0 \Big(
             {f^{q}_{1}}
          + 2  {B^{q}_{1}}
          - 2   {\gamma^{b}_{0}} \Big) 
          \Big)
       \Bigg]
\nonumber\\
& 
       + a_s^3 (q^2) \Bigg[
         \frac{1}{\epsilon^4}   \Big(
          - \frac{88}{9} \bt_0^2 {A^{q}_{1}}
          \Big)
       + \frac{1}{\epsilon^3}   \Big(
          - \frac{32}{9} \bt_1 {A^{q}_{1}}
          - \frac{20}{9} \bt_0 {A^{q}_{2}}
          + \frac{8}{3} \bt_0^2 \Big(
            {f^{q}_{1}}
          + 2  {B^{q}_{1}}
\nonumber\\
& 
          - 2 {\gamma^{b}_{0}} \Big)
          \Big)0
       + \frac{1}{\epsilon^2}   \Big(
          - \frac{4}{9} {A^{q}_{3}}
          + \frac{4}{3} \bt_1 \Big(
           {f^{q}_{1}}
          + 2 {B^{q}_{1}}
          - 2 {\gamma^{b}_{0}} \Big)
          + \frac{4}{3} \bt_0 \Big(
             {f^{q}_{2}}
          + 2  {B^{q}_{2}}
          - 2  {\gamma^{b}_{1}} \Big)
          \Big)
\nonumber\\
& 
       + \frac{1}{\epsilon}   \Big(
            \frac{2}{3} {f^{q}_{3}}
          + \frac{4}{3} {B^{q}_{3}}
          - \frac{4}{3} {\gamma^{b}_{2}}
          \Big)
       +  \Big(
            \frac{2}{3} {g^{b,1}_{3}}
          + \frac{4}{3} \bt_1 {g^{b,2}_{1}}
          + \frac{4}{3} \bt_0 {g^{b,2}_{2}}
          + \frac{8}{3} \bt_0^2 {g^{b,3}_{1}}
          + 3 \zeta_2 {A^{q}_{3}}
\nonumber\\
& 
          + 3 \zeta_2 \bt_1 \Big(
           {f^{q}_{1}}
          + 2   {B^{q}_{1}}
          - 2   {\gamma^{b}_{0}} \Big)
          + 6 \zeta_2 \bt_0 \Big(
            {f^{q}_{2}}
          + 2   {B^{q}_{2}}
          - 2   {\gamma^{b}_{1}} \Big)
\nonumber\\
&   
          - 12 \zeta_2 \bt_0^2 {g^{b,1}_{1}}
          - 3 \zeta_2^2 \bt_0^2 {A^{q}_{1}}
          \Big)
       \Bigg] \, .
\label{logF2}
\end{align}
Note that the poles of $\ln|{\hat F}^b|^2$ are fully controlled by the universal anomalous dimensions $A^q,\gamma^b,B^q$ and $f^q$ while the constant terms require vertex dependent constants $g_i^{b,k}$.
 
In $\overline{\rm{MS}}$ scheme, the mass factorization kernels $\Gamma_{bb}(\hat a_s,\mu^2,\mu_F^2,z_j,\ep)$ remove the collinear singularities which arise due to massless partons. These kernels satisfy the following RG equation :
\begin{equation}
 \mu_F^2 \frac{d}{d\mu_F^2} \Gamma_{bb}(z_j,\mu_F^2,\epsilon) = \frac{1}{2} \sum_c P_{bc} \left(z_j,\mu_F^2\right) \otimes \Gamma_{cb} \left(z_j,\mu_F^2,\epsilon \right) \, ,
 \label{rgeAP}
\end{equation}
where $P_{bc} \left(z_j,\mu_F^2\right)$ are AP splitting functions. We can expand the $P_{bc} \left(z_j,\mu_F^2\right)$ in powers of $a_s$ as 
\begin{eqnarray}
P_{bc}(z_j,\mu_F^2)=\sum_{i=1}^\infty a_s^i(\mu_F^2) P^{(i-1)}_{bc}(z_j).
\end{eqnarray}  
The off diagonal splitting functions are regular as $z_j \rightarrow 1$.  The diagonal ones contain in addition distributions such as $\delta(1-z_j)$ and ${\cal D}_0$ multiplied by the universal anomalous dimensions $B_{i}^q$ and $A_{i}^q$, respectively: 
\begin{equation}
P^{(i)}_{bb}(z_j) = 2 \Big( B_{i+1}^b \delta(1-z_j)+ A_{i+1}^b {\cal D}_0 \Big) + P_{reg,bb}^{(i)}(z_j) \,.
\end{equation}
As we are interested in results from the threshold region, we can ignore all the non-diagonal splitting functions and also the regular part $P_{reg,bb}^{(i)}$ arising from the diagonal terms.  Hence, the solution to Eq.~\ref{rgeAP} takes the following form:   
\begin{align}
\ln \Gamma_{bb}(z_{j},\mu_F^2)&=
         a_s(\mu_F^2) \Bigg[ 
          \delta(1-z_j) \left(  
          \frac{1}{\epsilon} \left(
           2 {B^{q}_{1}} \right)
          \right)
       + \D_0 \left(  \frac{1}{\epsilon} \left(  
            2 {A^{q}_{1}}
         \right) \right)
         \Bigg]
\nonumber\\
&
       +  a_s^2(\mu_F^2) \Bigg[ 
         \delta(1-z_j)  \left(
         \frac{1}{\epsilon^2}  \left(
            2 \bt_0 {B^{q}_{1}}
          \right)
       + \frac{1}{\epsilon}    \left(
            {B^{q}_{2}}
          \right)
        \right) 
       + \D_0  \left(
         \frac{1}{\epsilon^2}   \left(
            2 \bt_0 {A^{q}_{1}}
         \right)
       + \frac{1}{\epsilon}    \left(
            {A^{q}_{2}}
         \right)     
      \right) 
       \Bigg]
\nonumber\\
&
       + a_s^3(\mu_F^2) \Bigg[ 
         \delta(1-z_j)  \left(
         \frac{1}{\epsilon^3}    \left(
            \frac{8}{3} \bt_0^2 {B^{q}_{1}}
          \right)
       + \frac{1}{\epsilon^2}   \left(
            \frac{4}{3} \bt_1 {B^{q}_{1}}
          + \frac{4}{3} \bt_0 {B^{q}_{2}}
          \right)
       + \frac{1}{\epsilon}  \left(
            \frac{2}{3} {B^{q}_{3}}
          \right)
          \right) 
\nonumber\\
&
       + \D_0   \left(
           \frac{1}{\epsilon^3}  \left(
            \frac{8}{3} \bt_0^2 {A^{q}_{1}}
          \right)
       + \frac{1}{\epsilon^2}   \left(
            \frac{4}{3} \bt_1 {A^{q}_{1}}
          + \frac{4}{3} \bt_0 {A^{q}_{2}}
          \right)
       + \frac{1}{\epsilon}   \left(
            \frac{2}{3} {A^{q}_{3}}
         \right)
       \right)
       \Bigg] \, .
\label{logGbb}
\end{align}
Finally, we need to determine the soft distribution function $\Phi^{b}_d (\hat{a}_s, q^2, \mu^2, z_1, z_2, \epsilon)$ in $\Delta_{d, b}^{\rm SV}$.  Its most general form can be systematically constructed
if $\Phi^{b}_d$ also satisfies a differential equation similar to the  form factor.  It is indeed the case because the $q^2$ dependence and pole structure of $\Phi^b_d$ have to be similar to those of $\ln|{\hat F}^b|^2$ in order to obtain finite distribution $\Psi$ in the limit $\epsilon \rightarrow 0$ \cite{Ravindran:2005vv, Ravindran:2006cg}.  Hence, we propose that $\Phi^b_d$ satisfies 
\begin{equation}
 q^2 \frac{d}{dq^2} \Phi^b_d  = \frac{1}{2} \Big[ \overline K^b_d (\hat{a}_s, \frac{\mu_R^2}{\mu^2}, z_1, z_2,
\epsilon ) + \overline G^b_d (\hat{a}_s, \frac{q^2}{\mu_R^2}, \frac{\mu_R^2}{\mu^2}, z_1, z_2, \epsilon ) \Big] \, .
\label{sphi}
\end{equation}
It is natural to move all the singular terms in $\epsilon$ of $\Phi^b_d$ to  $\overline{K}^b_d$ and keep $\overline{G}^b_d$ finite as $\epsilon \rightarrow 0$ similar to $K^b_d$ and $G^b_d$ of the logarithm of the form factor, $\ln {\hat F}^b$. The RG invariance of $\Phi^{b}_d(\hat a_s,q^2,\mu^2,z_1,z_2,\ep)$ leads to 
\begin{equation}
\mu_R^2 \frac{d}{d\mu_R^2}\Phi^{b}_d(\hat a_s,q^2,\mu^2,z_1,z_2,\ep)=0\,
\end{equation}
and consequently 
\begin{equation}
\mu_R^2 \frac{d}{d\mu_R^2} \overline{K}^{b}_d = - \mu_R^2 \frac{d}{d\mu_R^2} \overline{G}^{b}_d = - \delta(1-z_1)\delta(1-z_2) a_s(\mu_R^2) \overline{A}^{q} \,.
\label{rgesKG}
\end{equation}
The right hand side of the above equation is proportion to $\delta(1-z_1) \delta(1-z_2)$ as the most singular terms resulting from $\overline K^b_d$ should cancel with those
from the form factor contribution which is proportional to only pure delta functions. To make the $\Delta_{d,b}^{\rm SV}$ finite, the poles from $\Phi^{b}_d (\hat{a}_s, q^2, \mu^2, z_{1}, z_{2}, \epsilon)$ have to cancel those coming from $\hat{F}^b$ and $\Gamma_{bb}$. Hence the constants $\overline{A}^{q}$ should satisfy 
\begin{equation}
 \overline{A}^{q} = - A^{q}\,.
\end{equation}
The RGE~\ref{rgesKG} for $\overline{G}^{b}_d$ can be solved using the above relation to get 
\begin{align}
& \overline G^{b}_d \left(\hat a_s, \frac{q^2}{\mu_R^2}, \frac{\mu_R^2}{\mu^2},z_1, z_2,\ep\right) 
\nonumber\\
& \qquad = \overline G^{b}_d \left( a_s(\mu_R^2), \frac{q^2}{\mu_R^2},z_1, z_2,\ep\right)
\nonumber\\
& \qquad = \overline G^{b}_d \left( a_s(q^2),1,z_1, z_2,\ep \right) - \delta(1-z_1) \delta(1-z_2)  \int_{ \frac{q^2}{\mu_R^2}}^1 \frac{d\lambda^2}{\lambda^2} A^q\left(a_s(\lambda^2 \mu_R^2)\right) \, .
\end{align}
With these solutions, it is now straightforward to solve the above differential equations \ref{sphi} for $\Phi^{b}_d$ to get 
\begin{align}
\Phi^{b}_d &= \Phi^{b}_d(\hat a_s,q^2 (1-z_1)(1-z_2),\mu^2,\ep)
\nonumber\\
& = \sum_{i=1}^\infty \hat{a}_s^i \left( \frac{q^2 (1-z_1)(1-z_2)}{\mu^2}\right)^{i \frac{\ep}{2}}\!\! S_{\ep}^i \left(\frac{(i~\ep)^2}{4(1-z_1) (1-z_2)} \right) \hat \phi^{b,(i)}_d(\ep)\,,
\label{softsol}
\end{align}
where, 
\begin{equation}
\hat \phi^{b,(i)}_d(\ep)= \frac{1}{i \ep} \Bigg[ \overline K^{b,(i)}_d(\ep) + \overline {G}^{b,(i)}_d(\ep)\Bigg]\,.
\end{equation}
The form of $z_j$ dependence part of the solution in the above solution is inspired by our one loop computation in the previous section and it can be justified from the factorization property of the QCD amplitudes and the corresponding partonic cross sections.  The constants $\overline K^{b,(i)}_d(\ep)$ are determined by expanding $\overline K^b_d$ in powers of $\hat a_s$ as follows
\begin{equation}
\overline K^b_d \left(\hat a_s, \frac{\mu_R^2}{\mu^2}, z_1, z_2, \ep \right) = \delta(1-z_1)\delta(1-z_2) \sum_{i=1}^\infty \hat a_s^i
   \left( \frac{\mu_R^2}{\mu^2} \right)^{i \frac{\ep}{2}}S^i_{\ep}~ \overline K^{b,(i)}_d(\ep) 
\label{Kbeps}
\end{equation}
and solving the RGE~\ref{rgesKG} for $\overline K^b_d$.   The constants $\overline K^{b,(i)}_d(\ep)$ are identical to $\overline K^{b,(i)}(\ep)$ given in \cite{Ravindran:2005vv, Ravindran:2006cg}. $\overline {G}^{b,(i)}_d(\ep)$ are related to the finite functions $\overline G^b_d(a_s(q^2),1,z_1,z_2,\ep)$. In terms of renormalized coupling constant, we find \begin{align}
\sum_{i=1}^\infty \hat{a}_s^i \left( \frac{q^2 (1-z_1)(1-z_2)}{\mu^2} \right)^{i \frac{\ep}{2}} S^i_{\ep}~ \overline G_d^{b,(i)}(\ep)
 = \sum_{i=1}^\infty a_s^i \left( q^2 (1-z_1)(1-z_2) \right) \overline{{\cal G}}^{b}_{d,i}(\ep) 
\label{Gbar1}
\end{align}
where the constants $\overline{{\cal G}}_{d,i}^b(\ep)$ are flavour independent and they satisfy the following structure similar to $G_i^b(\ep)$ of the form factor, i.e.,
\begin{align}
\overline {\cal G}^{b}_{d,i}(\ep) &=  - f_i^q + {\cal C}_{d,i}^b + \sum_{k=1}^\infty \ep^k \overline {\cal G}^{b,k}_{d,i}\,,
\label{OGI}
\end{align}
where
\begin{equation}
{\cal C}^b_{d,1} = 0,\quad \quad  {\cal C}^b_{d,2} = - 2 \beta_0 \overline {\cal G}_{d,1}^{b,1}, \quad \quad
% \nonumber\\[2ex]
{\cal C}^b_{d,3} = - 2 \beta_1 \overline {\cal G}_{d,1}^{b,1} - 2 \beta_0 ( 
\overline {\cal G}_{d,2}^{b,1} + 2 \beta_0  \overline {\cal G}_{d,1}^{b,2}) \, .
\end{equation}

Using $\overline K^{b,(i)}_d$ from Eq.~\ref{Kbeps} and  $\overline G^{b,(i)}_d$ from  Eq.~\ref{Gbar1} and using Eq.~\ref{Defplus}, we find that the soft distribution function
up to third order in $a_s(q^2)$ takes the form
\begin{align}
\Phi^b_d &=
%  asQ term
         a_s(q^2) \Bigg[ 
         \delta(1-z_1) \delta(1-z_2) \left(
          \frac{1}{\epsilon^2}   ( 2 {A^{q}_{1}}  )
       +  \frac{1}{\epsilon}  (
          - {f^{q}_{1}}
          )
       +   
           {\g^{q,1}_{d,1}}
       \right)
       + \D_0  \delta(1-z_2) 
       \Big(
       \frac{1}{\epsilon}   (
            {A^{q}_{1}}
          )
\nonumber\\
&
       +   (
          - \frac{1}{2} {f^{q}_{1}}
          ) 
       \Big)
       + \D_0 \DD_0  \left(
            \frac{1}{2} {A^{q}_{1}}
          \right)      
       + \D_1 \delta(1-z_2)    \left(
            \frac{1}{2} {A^{q}_{1}}
          \right)
       + \DD_0  \delta(1-z_1) 
       \left(
       \frac{1}{\epsilon}   (
            {A^{q}_{1}}
          )
       +  (
          - \frac{1}{2} {f^{q}_{1}}
          )
       \right)
\nonumber\\
&
       + \DD_1  \delta(1-z_1)  \left(   
            \frac{1}{2} {A^{q}_{1}}
         \right)  
       \Bigg]
% 
% %%%%%%%%% asQ^2 term 
       + a_s^2(q^2)  \Bigg[
       \delta(1-z_1) \delta(1-z_2)  \Bigg(
          \frac{1}{\epsilon^3}   \Big(
            3 \bt_0 {A^{q}_{1}}
          \Big)
       +  \frac{1}{\epsilon^2}   \Big(
            \frac{1}{2} {A^{q}_{2}}
          - \bt_0 {f^{q}_{1}}
          \Big)
\nonumber\\
&
       +  \frac{1}{\epsilon}   \Big(
          - \frac{1}{2} {f^{q}_{2}}
          \Big)
       +    \Big(
            \frac{1}{2} {\g^{q,1}_{d,2}}
          + \bt_0 {\g^{q,2}_{d,1}}
          \Big)
       \Bigg)
       + \D_0 \delta(1-z_2) \Bigg(
         \frac{1}{\epsilon^2}   \Big(
            \bt_0 {A^{q}_{1}}
          \Big)
       +  \frac{1}{\epsilon}  \Big(
            \frac{1}{2} {A^{q}_{2}}
          \Big)
\nonumber\\
&
       +   \Big(
          - \frac{1}{2} {f^{q}_{2}}
          - \bt_0 {\g^{q,1}_{d,1}}
          \Big)
       \Bigg)
       + \D_0 \DD_0    \Bigg(
            \frac{1}{2} {A^{q}_{2}}
          + \frac{1}{2} \bt_0 {f^{q}_{1}}
          \Bigg)       
       + \D_0 \DD_1   \Bigg(
          - \frac{1}{2} \bt_0 {A^{q}_{1}}
          \Bigg)       
\nonumber\\
&
       + \D_1  \delta(1-z_2)   \Bigg(
            \frac{1}{2} {A^{q}_{2}}
          + \frac{1}{2} \bt_0 {f^{q}_{1}}
          \Bigg)       
%        
% \nonumber\\
% &
       + \D_1 \DD_0   \Bigg(
          - \frac{1}{2} \bt_0 {A^{q}_{1}}
          \Bigg)   
       + \D_2  \delta(1-z_2)    \Bigg(
          - \frac{1}{4} \bt_0 {A^{q}_{1}}
          \Bigg)
\nonumber\\
&
       + \DD_0  \delta(1-z_1)   \Bigg(
         \frac{1}{\epsilon^2}    \Big(
            \bt_0 {A^{q}_{1}}
          \Big)
       +  \frac{1}{\epsilon}    \Big(
            \frac{1}{2} {A^{q}_{2}}
          \Big)
       +    \Big(
          - \frac{1}{2} {f^{q}_{2}}
          - \bt_0 {\g^{q,1}_{d,1}}
          \Big)
       \Bigg)
\nonumber\\
&
       + \DD_1  \delta(1-z_1)   \Bigg(
            \frac{1}{2} {A^{q}_{2}}
          + \frac{1}{2} \bt_0 {f^{q}_{1}}
          \Bigg)       
       + \DD_2  \delta(1-z_1)   \Bigg(
          - \frac{1}{4} \bt_0 {A^{q}_{1}}
          \Bigg)
       \Bigg]
\nonumber\\
&
% asQ^3 term
%        
       + a_s^3(q^2) \Bigg[
         \delta(1-z_1) \delta(1-z_2) \Bigg(
          \frac{1}{\epsilon^4}  \Big(
            \frac{44}{9} \bt_0^2 {A^{q}_{1}}
          \Big)
       +  \frac{1}{\epsilon^3}  \Big(
            \frac{16}{9} \bt_1 {A^{q}_{1}}
          + \frac{10}{9} \bt_0 {A^{q}_{2}}
          - \frac{4}{3} \bt_0^2 {f^{q}_{1}}
          \Big)
\nonumber\\
&
       +  \frac{1}{\epsilon^2}    \Big(
            \frac{2}{9} {A^{q}_{3}}
          - \frac{2}{3} \bt_1 {f^{q}_{1}}
          - \frac{2}{3} \bt_0 {f^{q}_{2}}
          \Big)
       -  \frac{1}{\epsilon} \Big(
           \frac{1}{3} {f^{q}_{3}}
          \Big)
       +  \Big(
            \frac{1}{3} {\g^{q,1}_{d,3}}
          + \frac{2}{3} \bt_1 {\g^{q,2}_{d,1}}
          + \frac{2}{3} \bt_0 {\g^{q,2}_{d,2}}
          + \frac{4}{3} \bt_0^2 {\g^{q,3}_{d,1}}
          \Big)
       \Bigg)
\nonumber\\
&
       + \D_0  \delta(1-z_2)  \Bigg(
         \frac{1}{\epsilon^3}   \Big(
            \frac{4}{3} \bt_0^2 {A^{q}_{1}}
          \Big)
       + \frac{1}{\epsilon^2}    \Big(
            \frac{2}{3} \bt_1 {A^{q}_{1}}
          + \frac{2}{3} \bt_0 {A^{q}_{2}}
          \Big)
       + \frac{1}{\epsilon}    \Big(
            \frac{1}{3} {A^{q}_{3}}
          \Big)
       +   \Big(
          - \frac{1}{2} {f^{q}_{3}}
          - \bt_1 {\g^{q,1}_{d,1}}
\nonumber\\
&
          - \bt_0 {\g^{q,1}_{d,2}}
          - 2 \bt_0^2 {\g^{q,2}_{d,1}}
          \Big)
       \Bigg)
       + \D_0 \DD_0    \Bigg(
            \frac{1}{2} {A^{q}_{3}}
          + \frac{1}{2} \bt_1 {f^{q}_{1}}
          + \bt_0 {f^{q}_{2}}
          + 2 \bt_0^2 {\g^{q,1}_{d,1}}
          \Bigg)
       + \D_0 \DD_1    \Bigg(
          - \frac{1}{2} \bt_1 {A^{q}_{1}}
\nonumber\\
&
          - \bt_0 {A^{q}_{2}}
          - \bt_0^2 {f^{q}_{1}}
          \Bigg)
       + \D_0 \DD_2   \Bigg(
            \frac{1}{2} \bt_0^2 {A^{q}_{1}}
          \Bigg)
       + \D_1  \delta(1-z_2)    \Bigg(
            \frac{1}{2} {A^{q}_{3}}
          + \frac{1}{2} \bt_1 {f^{q}_{1}}
          + \bt_0 {f^{q}_{2}}
          + 2 \bt_0^2 {\g^{q,1}_{d,1}}
          \Bigg)
\nonumber\\
&
       + \D_1 \DD_0   \Bigg(
          - \frac{1}{2} \bt_1 {A^{q}_{1}}
          - \bt_0 {A^{q}_{2}}
          - \bt_0^2 {f^{q}_{1}}
          \Bigg)
       + \D_1 \DD_1   \Bigg(
            \bt_0^2 {A^{q}_{1}}
          \Bigg)
\nonumber\\
&
       + \D_2  \delta(1-z_2)   \Bigg(
          - \frac{1}{4} \bt_1 {A^{q}_{1}}
          - \frac{1}{2} \bt_0 {A^{q}_{2}}
% \nonumber\\
% &
          - \frac{1}{2} \bt_0^2 {f^{q}_{1}}
          \Bigg)
       + \D_2 \DD_0    \Bigg(
            \frac{1}{2} \bt_0^2 {A^{q}_{1}}
          \Bigg)
\nonumber\\
&
       + \D_3   \delta(1-z_2)   \Bigg(
            \frac{1}{6} \bt_0^2 {A^{q}_{1}}
          \Bigg)
       +  \DD_0  \delta(1-z_1)   \Bigg(  
         \frac{1}{\epsilon^3}  \Big(
            \frac{4}{3} \bt_0^2 {A^{q}_{1}}
          \Big)
% 
% \nonumber\\
% &
       + \frac{1}{\epsilon^2}    \Big(
            \frac{2}{3} \bt_1 {A^{q}_{1}}
          + \frac{2}{3} \bt_0 {A^{q}_{2}}
          \Big)
\nonumber\\
&
       + \frac{1}{\epsilon}    \Big(
            \frac{1}{3} {A^{q}_{3}}
          \Big)
       +    \Big(
          - \frac{1}{2} {f^{q}_{3}}
          - \bt_1 {\g^{q,1}_{d,1}}
          - \bt_0 {\g^{q,1}_{d,2}}
          - 2 \bt_0^2 {\g^{q,2}_{d,1}}
          \Big)
       \Bigg)
\nonumber\\
&
       + \DD_1  \delta(1-z_1)    \Bigg(
            \frac{1}{2} {A^{q}_{3}}
          + \frac{1}{2} \bt_1 {f^{q}_{1}}
          + \bt_0 {f^{q}_{2}}
          + 2 \bt_0^2 {\g^{q,1}_{d,1}}
          \Bigg)
\nonumber\\
&
       + \DD_2 \delta(1-z_1)   \Bigg(
          - \frac{1}{4} \bt_1 {A^{q}_{1}}
          - \frac{1}{2} \bt_0 {A^{q}_{2}}
          - \frac{1}{2} \bt_0^2 {f^{q}_{1}}
          \Bigg)
\nonumber\\
&
       + \DD_3  \delta(1-z_1)   \Bigg(
            \frac{1}{6} \bt_0^2 {A^{q}_{1}}
          \Bigg)
       \Bigg] \, .
\label{Phidb}
\end{align}
In the above expression, we have used ${\overline {\cal G}}_{d,i}^{b,k} = {\overline {\cal G}}_{d,i}^{q,k}$, being flavour independent. The soft distribution function depends in addition to the universal anomalous dimensions $A_i^q$,$B_i^q$,$\gamma_i^q$ and $f_i^q$, the constants $\overline{\cal G}^{q,k}_{d,i}$ which need to to be determined. At ${\cal O}(a_s)$ level $\overline{\cal G}^{q,1}_{d,1}, \, \overline{\cal G}^{q,1}_{d,2}, \, \overline{\cal G}^{q,1}_{d,3}$, at ${\cal O}(a_s^2)$  $\overline{\cal G}^{q,2}_{d,1}$,  $\overline{\cal G}^{q,2}_{d,2}$ and at ${\cal O}(a_s^3)$ $\overline{\cal G}^{q,3}_{d,1}$ are needed to obtain $\Phi^b_d$.  We achieve this using the following identity:
\begin{equation}
\int_0^1 dx_1^0 \int_0^1 dx_2^0 \left(x_1^0 x_2^0\right)^{N-1} \frac{d \sigma^b}{d y}
=\int_0^1 d\tau~ \tau^{N-1} ~\sigma^b\,,
\label{iden}
\end{equation}
where $\sigma^b$ is known to NNLO level \cite{Harlander:2003ai} exactly and to N${}^3$LO level in the threshold limit \cite{Ahmed:2014cha}. In large N limit \textit{i.e.} $N \rightarrow \infty$ the above Eq.~\ref{iden} relates $\hat{\phi}^{q,(i)}_d(\ep)$ to $\hat{\phi}^{q,(i)} (\ep)$ that appears in inclusive threshold corrections to Drell-Yan process (see \cite{Ravindran:2005vv, Ravindran:2006cg, Ahmed:2014cla, Ahmed:2014cha}) as follows 
\begin{align}
&\hat{\phi}^{b,(i)}_d(\ep) = \frac{\Gamma(1+i~\ep)}{\Gamma^2\left(1+i \frac{\ep}{2}\right)} \hat{\phi}^{b,(i)}(\ep) \\
\intertext{and}
&\hat \phi^{b,(i)}(\ep) = \hat \phi^{q,(i)}(\ep) 
\end{align}
since $\hat{\phi}^{q,(i)}(\ep)$ is flavour independent.  Hence
\begin{eqnarray}
\hat{\phi}^{b,(i)}_d(\ep)=\hat{\phi}^{q,(i)}_d(\ep)
\end{eqnarray}
and all the relevant constants $\overline {\cal G}_{d,i}^{q,k}$ required for threshold prediction up to ${\cal O}(a_s^3)$ can be obtained from $\overline{\cal G}_{i}^{q,k}$ which are analogous to these factors appeared in the computation of inclusive threshold cross-section to Drell-Yan process. The relevant $\overline{\cal G}_{i}^{q,k}$'s  at ${\cal O}(a_{s})$ and ${\cal O}(a_{s}^{2})$ \cite{Ravindran:2005vv, Ravindran:2006cg} are
\begin{align*}
  {\overline {\cal G}}^{q,1}_1 &= C_F ( - 3 \zeta_2 ) \,, \\
  {\overline {\cal G}}^{q,2}_1 &= C_F ( \frac{7}{3} \zeta_3 ) \,, \\
  {\overline {\cal G}}^{q,3}_1 &=  C_F ( - \frac{3}{16} {\zeta_2}^2 ) \,, \\
  {\overline {\cal G}}^{q,1}_2 &=  C_F n_f  \Big( - \frac{328}{81} + \frac{70}{9} \zeta_2 + \frac{32}{3} \zeta_3 \Big)
             + C_A C_F  \Big( \frac{2428}{81} - \frac{469}{9} \zeta_2 
                       + 4 {\zeta_2}^2 - \frac{176}{3} \zeta_3 \Big) \,, \\
  {\overline {\cal G}}^{q,2}_2 &=  C_A C_F \Big( \frac{11}{40} {\zeta_2}^2 - \frac{203}{3} {\zeta_2} {\zeta_3}
             + \frac{1414}{27} {\zeta_2} + \frac{2077}{27} {\zeta_3} + 43 {\zeta_5} - \frac{7288}{243}  \Big) \\
           & + C_F n_f \Big( -\frac{1}{20} {\zeta_2}^2 - \frac{196}{27} {\zeta_2} - \frac{310}{27} {\zeta_3} + \frac{976}{243} \Big) \\
\end{align*}
and at ${\cal O}(a_{s}^{3})$ \cite{Ahmed:2014cla} 
\begin{align}
% 
%%%%%%%%%%%%%%% gbar31b %%%
{\overline {\cal G}}^{q,1}_3 &= 
C_F \Big\{  {C_A}^2 \Big(\frac{152}{63} \;{\zeta_2}^3 + \frac{1964}{9} \;{\zeta_2}^2
+ \frac{11000}{9} \;{\zeta_2} {\zeta_3} - \frac{765127}{486} \;{\zeta_2}
+\frac{536}{3} \;{\zeta_3}^2 - \frac{59648}{27} \;{\zeta_3} 
\nonumber\\
&
- \frac{1430}{3} \;{\zeta_5}
+\frac{7135981}{8748}\Big)
+{C_A} {n_f} \
\Big(-\frac{532}{9} \;{\zeta_2}^2 - \frac{1208}{9} \;{\zeta_2} {\zeta_3}
+\frac{105059}{243} \;{\zeta_2} + \frac{45956}{81} \;{\zeta_3} 
\nonumber\\
&
+\frac{148}{3} \;{\zeta_5} - \frac{716509}{4374} \Big)
+ {C_F} {n_f} \
\Big(\frac{152}{15} \;{\zeta_2}^2 
- 88 \;{\zeta_2} {\zeta_3} 
+\frac{605}{6} \;{\zeta_2} + \frac{2536}{27} \;{\zeta_3}
+\frac{112}{3} \;{\zeta_5} 
\nonumber\\
&
- \frac{42727}{324}\Big)
+ {n_f}^2 \
\Big(\frac{32}{9} \;{\zeta_2}^2 - \frac{1996}{81} \;{\zeta_2}
-\frac{2720}{81} \;{\zeta_3} + \frac{11584}{2187}\Big)  \Big\} \,.
\end{align}
These lead to the following expressions of $\overline {\cal G}_{d,i}^{q,k}$'s at ${\cal O}(a_{s})$, ${\cal O}(a_{s}^{2})$ \cite{Ravindran:2006bu}  and ${\cal O}(a_{s}^{3})$ \cite{Ahmed:2014uya} :
\begin{align}
\label{eq:calgd}
\overline {\cal G}_{d,1}^{q,1} &= - C_{F} \zeta_{2} \,, \quad\quad
\overline {\cal G}_{d,2}^{q,1} = C_{F} \Bigg\{ \frac{1}{3} \zeta_{3} \Bigg\} \,, \quad\quad
\overline {\cal G}_{d,3}^{q,1} = C_{F} \Bigg\{ \frac{1}{80} \zeta_{2}^{2} \Bigg\} \,,
\nonumber\\
\overline {\cal G}_{d,1}^{q,2} &= {C_A} {C_F} \Bigg\{ -4 {\zeta_2}^2-\frac{67 \
                                }{3} {\zeta_2} -\frac{44 }{3} {\zeta_3} + \frac{2428}{81} \Bigg\} 
                                + C_{F} n_{f} \Bigg\{ \frac{8}{3} \zeta_{3} + \frac{10}{3} \zeta_{2} - \frac{328}{81} \Bigg\} \,,
\nonumber\\
\overline {\cal G}_{d,2}^{q,2} &= {C_A} {C_F} \Bigg\{ -\frac{319 }{120} {\zeta_2}^2 - \frac{71 \
                                }{3} {\zeta_2} {\zeta_3} + \frac{202 }{9} {\zeta_2} + \frac{469 \
                                }{27} {\zeta_3} + 43 {\zeta_5}-\frac{7288}{243} \Bigg\} 
\nonumber\\
                             &  + {C_F} {n_f} \Bigg\{ \frac{29 }{60} {\zeta_2}^2 - \frac{28 \
                                }{9} {\zeta_2} - \frac{70 }{27} {\zeta_3} + \frac{976}{243} \Bigg\} \,,
\nonumber\\%
\overline {\cal G}_{d,1}^{q,3} &= {C_A}^2 {C_F} \Bigg\{ \frac{17392 }{315} {\zeta_2}^3 + \frac{1538 \
                                }{45} {\zeta_2}^2 + \frac{4136 }{9} {\zeta_2} {\zeta_3} - \frac{379417 \
                                }{486} {\zeta_2} + \frac{536 }{3} {\zeta_3}^2 
                                - 936 {\zeta_3} 
\nonumber\\ 
                             & - \frac{1430 \
                               }{3} {\zeta_5} + \frac{7135981}{8748} \Bigg\} 
                             + {C_A} {C_F} {n_f} \Bigg\{ -\frac{1372 \
                               }{45} {\zeta_2}^2 -\frac{392}{9} {\zeta_2} {\zeta_3} + \frac{51053 \
                               }{243} {\zeta_2} 
\nonumber\\
                             & + \frac{12356}{81} {\zeta_3} + \frac{148 \ 
                               }{3} {\zeta_5} - \frac{716509}{4374} \Bigg\}
                             + {C_F} {n_f}^2 \Bigg\{ \frac{152}{45} {\zeta_2}^2 - \frac{316 \
                               }{27} {\zeta_2} - \frac{320 }{81} {\zeta_3} + \frac{11584}{2187} \Bigg\}
\nonumber\\
                            & + {C_F}^2 {n_f} \Bigg\{ \frac{152}{15} {\zeta_2}^2 - 40 {\zeta_2} \
                                {\zeta_3}+\frac{275 }{6} {\zeta_2} + \frac{1672 \
                                }{27} {\zeta_3} + \frac{112 }{3} {\zeta_5} - \frac{42727}{324} \Bigg\} \, .
\end{align}
With all these information available at hand, it is now straightforward to obtain threshold corrections to rapidity distribution of Higgs boson in the bottom quark annihilation processes.  We substitute Eq.~\ref{logZb}, \ref{logF2}, \ref{logGbb}, \ref{Phidb} in Eq.~\ref{DYH} to obtain $\Psi^b_d(\ep)$.  Since all the UV and IR singularities cancel among various terms, we can set $\ep=0$ in the the distribution $\Psi^b_d(\ep)$ to obtain $\Delta^{\rm SV}_{d,b}$.   Expanding the finite distribution $\Psi^b_d(\ep=0)$ in Eq.~\ref{master} in terms of convolutions Eq.~\ref{convexp} and performing all those convolutions using the formula given in Eq.~52 of \cite{Ravindran:2006cg}, we obtain $\Delta^{{\rm SV},(i)}_{b}$ defined by
\begin{eqnarray}
\Delta^{\rm SV}_{d,b}(z_1,z_2,q^2,\mu_R^2,\mu_F^2)
=\sum_{i=0}^\infty a_s^{i}(q^2) \Delta^{{\rm SV},(i)}_{d,b}(z_1,z_2,q^2,\mu_R^2,\mu_F^2) 
\end{eqnarray}
We present below our results for $\Delta^{{\rm SV},(i)}_{d,b}$ up to N$^3$LO level in terms of  of the constants $A^q_j$, $B^q_j$, $f^q_j$, $\gamma^b_j$, $\beta_j$, $g^{b,k}_{j}$ and $\overline{{\cal G}}_{d,j}^{q,k}$:
\begin{align}
&\Delta_{d,b}^{{\rm SV},(1)} =
        \delta(1-z_1) \delta(1-z_2)   \Big[
             \g^{q,1}_{d,1}
          +  g^{b,1}_{1}
          + \frac{3}{2} \zeta_2 A^{q}_{1}
          \Big]
       + \D_0 \delta(1-z_2)   \Big[
          - f^{q}_{1}
          \Big]
\nonumber \\
& \quad
       + \D_0 \DD_0   \Big[
           \frac{1}{2} A^{q}_{1}
          \Big]
       + \D_1 \delta(1-z_2)   \Big[
           A^{q}_{1}
          \Big]
       + \Bigg\{ z_1 \leftrightarrow z_2 \Bigg\}
\nonumber \\
%
%%%%%%%%%%%%%%%%%%%%%%%%%%%%%%%%%%%%%%%%%% CI2
% 
 &\Delta_{d,b}^{{\rm SV},(2)} =
         \delta(1-z_1) \delta(1-z_2)   \Bigg[
            \frac{1}{2} \g^{q,1}_{d,2}
          + {\g^{q,1}_{d,1}}^2
          + \frac{1}{2} g^{b,1}_{2}
          + 2 g^{b,1}_{1} \g^{q,1}_{d,1}
          + {g^{b,1}_{1}}^2
          + \bt_0 \Big(
            \g^{q,2}_{d,1}
          + g^{b,2}_{1}
          \Big)
\nonumber \\
& \quad
          - \zeta_3 A^{q}_{1} f^{q}_{1}
          + \zeta_2 \Big(
          - \frac{1}{2} ({f^{q}_{1}})^2
          + \frac{3}{2} A^{q}_{2}
          + 3  \g^{q,1}_{d,1} A^{q}_{1}
          + 3  g^{b,1}_{1} A^{q}_{1}
          \Big)
          + \zeta_2 \bt_0 \Big(
            \frac{3}{2}  f^{q}_{1}
          + 3  B^{q}_{1}
          - 3  \gamma^{b}_0
          \Big)
\nonumber \\
& \quad
          + \frac{49}{20} \zeta_2^2 ({A^{q}_{1}})^2
          \Bigg]
       + \D_0 \delta(1-z_2)   \Bigg[
          - f^{q}_{2}
          - 2 \g^{q,1}_{d,1} f^{q}_{1}
          - 2 g^{b,1}_{1} f^{q}_{1}
          - 2 \bt_0 \g^{q,1}_{d,1}
          + 2 \zeta_3 ({A^{q}_{1}})^2
\nonumber \\
& \quad
          - \zeta_2 A^{q}_{1} f^{q}_{1}
          \Bigg]
       + \D_0 \DD_0   \Bigg[
            \frac{1}{2} ({f^{q}_{1}})^2
          + \frac{1}{2} A^{q}_{2}
          + \g^{q,1}_{d,1} A^{q}_{1}
          + g^{b,1}_{1} A^{q}_{1}
          + \frac{1}{2} \bt_0 f^{q}_{1}
          + \frac{1}{2} \zeta_2 ({A^{q}_{1}})^2
          \Bigg]
\nonumber \\
& \quad
       + \D_1 \delta(1-z_2)   \Bigg[
            ({f^{q}_{1}})^2
          + A^{q}_{2}
          + 2 \g^{q,1}_{d,1} A^{q}_{1}
          + 2 g^{b,1}_{1} A^{q}_{1}
          + \bt_0 f^{q}_{1}
          + \zeta_2 ({A^{q}_{1}})^2
          \Bigg]
\nonumber \\
& \quad
       + \D_1 \DD_0   \Bigg[
          - 3 A^{q}_{1} f^{q}_{1}
          - \bt_0 A^{q}_{1}
          \Bigg]
       + \D_1 \DD_1   \Bigg[
           \frac{3}{2} ({A^{q}_{1}})^2
          \Bigg]
       + \D_2 \delta(1-z_2)   \Bigg[
          - \frac{3}{2} A^{q}_{1} f^{q}_{1}
          - \frac{1}{2} \bt_0 A^{q}_{1}
          \Bigg]
\nonumber \\
& \quad
       + \D_2 \DD_0   \Bigg[
            \frac{3}{2} ({A^{q}_{1}})^2
          \Bigg]
       + \D_3 \delta(1-z_2)   \Bigg[
            \frac{1}{2} ({A^{q}_{1}})^2
          \Bigg]
       + \Bigg\{ z_1 \leftrightarrow z_2 \Bigg\}
\end{align}

%%%%%%%%%%%%%%%%%% general CI3
\begin{align}
&\Delta_{d,b}^{{\rm SV},(3)} =
        \delta(1-z_1) \delta(1-z_2)   \Bigg[
            \frac{1}{3} {\g^{q,1}_{d,3}} 
          + {\g^{q,1}_{d,1}} {\g^{q,1}_{d,2}}
          + \frac{2}{3} {\g^{q,1}_{d,1}}^3
          + \frac{1}{3} {g^{b,1}_{3}}
          + {g^{b,1}_{2}} {\g^{q,1}_{d,1}}
          + {g^{b,1}_{1}} {\g^{q,1}_{d,2}}
\nonumber \\
& \quad
          + 2 {g^{b,1}_{1}} {\g^{q,1}_{d,1}}^2
          + {g^{b,1}_{1}} {g^{b,1}_{2}}
          + 2 {g^{b,1}_{1}}^2 {\g^{q,1}_{d,1}}
          + \frac{2}{3} {g^{b,1}_{1}}^3
          + \frac{2}{3} \bt_1 \Big( {\g^{q,2}_{d,1}}
                                  + {g^{b,2}_{1}} \Big)
          + 2 \bt_0 \Big( \frac{1}{3}  {\g^{q,2}_{d,2}}
          + {\g^{q,1}_{d,1}} {\g^{q,2}_{d,1}}
\nonumber \\
& \quad
          + \frac{1}{3} {g^{b,2}_{2}}
          + {g^{b,2}_{1}} {\g^{q,1}_{d,1}}
          + {g^{b,1}_{1}} {\g^{q,2}_{d,1}}
          + {g^{b,1}_{1}} {g^{b,2}_{1}}
          \Big)
          + \frac{4}{3} \bt_0^2 \Big( {\g^{q,3}_{d,1}} + {g^{b,3}_{1}} \Big)
          - 3 \zeta_5 ({A^{q}_{1}})^2 {f^{q}_{1}}
          - 2 \zeta_5 \bt_0 ({A^{q}_{1}})^2
\nonumber \\
& \quad
          - \zeta_3 \Big(
           \frac{1}{3} ({f^{q}_{1}})^3
          + {A^{q}_{2}} {f^{q}_{1}}
          + {A^{q}_{1}} {f^{q}_{2}}
          + 2 {\g^{q,1}_{d,1}} {A^{q}_{1}} {f^{q}_{1}}
          + 2 {g^{b,1}_{1}} {A^{q}_{1}} {f^{q}_{1}}
          + \bt_0 ({f^{q}_{1}})^2
          + 2 \bt_0 {\g^{q,1}_{d,1}} {A^{q}_{1}}
          \Big)
\nonumber \\
& \quad
          + \frac{5}{3} \zeta_3^2 ({A^{q}_{1}})^3
          + \zeta_2 \Big(
          - {f^{q}_{1}} {f^{q}_{2}}
          + \frac{3}{2} {A^{q}_{3}}
          + \frac{3}{2} {\g^{q,1}_{d,2}} {A^{q}_{1}}
          - {\g^{q,1}_{d,1}} ({f^{q}_{1}})^2
          + 3 {\g^{q,1}_{d,1}} {A^{q}_{2}}
          + 3 {\g^{q,1}_{d,1}}^2 {A^{q}_{1}}
\nonumber \\
& \quad
          + \frac{3}{2} {g^{b,1}_{2}} {A^{q}_{1}}
          - {g^{b,1}_{1}} ({f^{q}_{1}})^2
          + 3 {g^{b,1}_{1}} {A^{q}_{2}}
          + 6 {g^{b,1}_{1}} {\g^{q,1}_{d,1}} {A^{q}_{1}}
          + 3  {g^{b,1}_{1}}^2 {A^{q}_{1}}
          \Big)
          + \frac{3}{2} \zeta_2 \bt_1 \Big(
            {f^{q}_{1}}
            + 2 {B^{q}_{1}}
          - 2 {\gamma^{b}_{0}}
          \Big)
\nonumber \\
& \quad
          + 3 \zeta_2 \bt_0 \Big(
            {f^{q}_{2}}
          + 2 {B^{q}_{2}}
          + {\g^{q,2}_{d,1}} {A^{q}_{1}}
          + \frac{1}{3} {\g^{q,1}_{d,1}} {f^{q}_{1}}
          + 2 {\g^{q,1}_{d,1}} {B^{q}_{1}}
          + {g^{b,2}_{1}} {A^{q}_{1}}
          + {g^{b,1}_{1}} {f^{q}_{1}}
          + 2 {g^{b,1}_{1}} {B^{q}_{1}}
          - 2 {\gamma^{b}_{1}}
\nonumber \\
& \quad
          - 2 {\gamma^{b}_{0}} {\g^{q,1}_{d,1}}
          - 2 {\gamma^{b}_{0}} {g^{b,1}_{1}}
          \Big)
          - 6 \zeta_2 \bt_0^2 {g^{b,1}_{1}}
          + \zeta_2 \zeta_3 ({A^{q}_{1}})^2 {f^{q}_{1}}
          + 2 \zeta_2 \zeta_3 \bt_0 ({A^{q}_{1}})^2
          + \frac{49}{10} \zeta_2^2 \Big(
          - \frac{11}{49} {A^{q}_{1}} ({f^{q}_{1}})^2
\nonumber \\
& \quad
          +  {A^{q}_{1}} {A^{q}_{2}}
          +  {\g^{q,1}_{d,1}} ({A^{q}_{1}})^2
          +  {g^{b,1}_{1}} ({A^{q}_{1}})^2
          \Big)
          + \frac{9}{2} \zeta_2^2 \bt_0 \Big( 
            {A^{q}_{1}} {f^{q}_{1}}
          + 2 {A^{q}_{1}} {B^{q}_{1}}
          - 2 {\gamma^{b}_{0}} {A^{q}_{1}}
          - \frac{1}{3} {A^{q}_{1}}
          \Big)
\nonumber \\
& \quad
          + \frac{1181}{420} \zeta_2^3 ({A^{q}_{1}})^3
          \Bigg]
% 
%%%%%%%%%%%%%%%%%%% D0 delta(1-z2) terms
% 
       + \D_0 \delta(1-z_2)   \Bigg[
          - {f^{q}_{3}}
          - {\g^{q,1}_{d,2}} {f^{q}_{1}}
          - 2 {\g^{q,1}_{d,1}} {f^{q}_{2}}
          - 2 {\g^{q,1}_{d,1}}^2 {f^{q}_{1}}
          - {g^{b,1}_{2}} {f^{q}_{1}}
          - 2 {g^{b,1}_{1}} {f^{q}_{2}}
\nonumber \\
& \quad
          - 4 {g^{b,1}_{1}} {\g^{q,1}_{d,1}} {f^{q}_{1}}
          - 2 {g^{b,1}_{1}}^2 {f^{q}_{1}}
          - 2 \bt_1 {\g^{q,1}_{d,1}}
          - 2 \bt_0 \Big(
            {\g^{q,1}_{d,2}}
          + {\g^{q,2}_{d,1}} {f^{q}_{1}}
          + 2 {\g^{q,1}_{d,1}}^2
          + {g^{b,2}_{1}} {f^{q}_{1}}
          + 2 {g^{b,1}_{1}} {\g^{q,1}_{d,1}}
          \Big)
\nonumber \\
& \quad
          - 4 \bt_0^2 {\g^{q,2}_{d,1}}
          + 6 \zeta_5 ({A^{q}_{1}})^3
          + 4 \zeta_3 \Big(
            {A^{q}_{1}} ({f^{q}_{1}})^2
          + {A^{q}_{1}} {A^{q}_{2}}
          + {\g^{q,1}_{d,1}} ({A^{q}_{1}})^2
          + {g^{b,1}_{1}} ({A^{q}_{1}})^2
          + \frac{3}{2} \bt_0 {A^{q}_{1}} {f^{q}_{1}}
          \Big)
\nonumber \\
& \quad
          + \zeta_2 \Big(
            ({f^{q}_{1}})^3
          - {A^{q}_{2}} {f^{q}_{1}}
          - {A^{q}_{1}} {f^{q}_{2}}
          - 2 {\g^{q,1}_{d,1}} {A^{q}_{1}} {f^{q}_{1}}
          - 2 {g^{b,1}_{1}} {A^{q}_{1}} {f^{q}_{1}}
          \Big)
          - \zeta_2 \bt_0 \Big(
            ({f^{q}_{1}})^2
          + 6 {B^{q}_{1}} {f^{q}_{1}}
          + 2 {\g^{q,1}_{d,1}} {A^{q}_{1}}
\nonumber \\
& \quad
          - 6 {\gamma^{b}_{0}} {f^{q}_{1}}
          \Big)
          - 2 \zeta_2 \zeta_3 ({A^{q}_{1}})^3
          - \frac{1}{2} \zeta_2^2 ({A^{q}_{1}})^2 {f^{q}_{1}}
          \Bigg]
%           
%%%%%%%%%%%% D0 DD0 term         
%           
       + \D_0 \DD_0   \Bigg[
            {f^{q}_{1}} {f^{q}_{2}}
          + \frac{1}{2} {A^{q}_{3}}
          + \frac{1}{2} {\g^{q,1}_{d,2}} {A^{q}_{1}}
          + {\g^{q,1}_{d,1}} ({f^{q}_{1}})^2
\nonumber \\
& \quad
          + {\g^{q,1}_{d,1}} {A^{q}_{2}}
          + {\g^{q,1}_{d,1}}^2 {A^{q}_{1}}
          + \frac{1}{2} {g^{b,1}_{2}} {A^{q}_{1}}
          + {g^{b,1}_{1}} ({f^{q}_{1}})^2
          + {g^{b,1}_{1}} {A^{q}_{2}}
          + 2 {g^{b,1}_{1}} {\g^{q,1}_{d,1}} {A^{q}_{1}}
          + {g^{b,1}_{1}}^2 {A^{q}_{1}}
          + \frac{1}{2} \bt_1 {f^{q}_{1}}
\nonumber \\
& \quad
          + \bt_0 \Big(
            {f^{q}_{2}}
          + {\g^{q,2}_{d,1}} {A^{q}_{1}}
          + 3 {\g^{q,1}_{d,1}} {f^{q}_{1}}
          + {g^{b,2}_{1}} {A^{q}_{1}}
          + {g^{b,1}_{1}} {f^{q}_{1}}
          \Big)
          + 2 \bt_0^2 {\g^{q,1}_{d,1}}
          - 5 \zeta_3 ({A^{q}_{1}})^2 {f^{q}_{1}}
          - 3 \zeta_3 \bt_0 ({A^{q}_{1}})^2
\nonumber \\
& \quad
          + \zeta_2 \Big(
            {A^{q}_{1}} {A^{q}_{2}}          
          - {A^{q}_{1}} ({f^{q}_{1}})^2
          + {\g^{q,1}_{d,1}} ({A^{q}_{1}})^2
          + {g^{b,1}_{1}} ({A^{q}_{1}})^2
          \Big)
          + 3 \zeta_2 \bt_0 \Big(
            {A^{q}_{1}} {B^{q}_{1}}
          - {\gamma^{b}_{0}} {A^{q}_{1}}
          \Big)
          + \frac{1}{4} \zeta_2^2 ({A^{q}_{1}})^3
          \Bigg]
\nonumber \\
& \quad
% 
%%%%%%%%%% D1 delta(1-z2) term %%%
%
       + \D_1 \delta(1-z_2)  \Bigg[
            2 {f^{q}_{1}} {f^{q}_{2}}
          + {A^{q}_{3}}
          + {\g^{q,1}_{d,2}} {A^{q}_{1}}
          + 2 {\g^{q,1}_{d,1}} ({f^{q}_{1}})^2
          + 2 {\g^{q,1}_{d,1}} {A^{q}_{2}}
          + 2 {\g^{q,1}_{d,1}}^2 {A^{q}_{1}}
          + {g^{b,1}_{2}} {A^{q}_{1}}
\nonumber \\
& \quad
          + 2 {g^{b,1}_{1}} ({f^{q}_{1}})^2
          + 2 {g^{b,1}_{1}} {A^{q}_{2}}
          + 4 {g^{b,1}_{1}} {\g^{q,1}_{d,1}} {A^{q}_{1}}
          + 2 {g^{b,1}_{1}}^2 {A^{q}_{1}}
          + \bt_1 {f^{q}_{1}}
          + 2 \bt_0 \Big(
            {f^{q}_{2}}
          + {\g^{q,2}_{d,1}} {A^{q}_{1}}
          + 3 {\g^{q,1}_{d,1}} {f^{q}_{1}}
\nonumber \\
& \quad
          + {g^{b,2}_{1}} {A^{q}_{1}}
          + {g^{b,1}_{1}} {f^{q}_{1}}
          \Big)
          + 4 \bt_0^2 {\g^{q,1}_{d,1}}
          - 10 \zeta_3 ({A^{q}_{1}})^2 {f^{q}_{1}}
          - 6 \zeta_3 \bt_0 ({A^{q}_{1}})^2
          + 2 \zeta_2 \Big(
          - {A^{q}_{1}} ({f^{q}_{1}})^2
          + {A^{q}_{1}} {A^{q}_{2}}
\nonumber \\
& \quad
          + {\g^{q,1}_{d,1}} ({A^{q}_{1}})^2
          + {g^{b,1}_{1}} ({A^{q}_{1}})^2
          \Big)
          + 6 \zeta_2 \bt_0 \Big(
            {A^{q}_{1}} {B^{q}_{1}}
          - {\gamma^{b}_{0}} {A^{q}_{1}}
          \Big)
          + \frac{1}{2} \zeta_2^2 ({A^{q}_{1}})^3
          \Bigg]
% 
%%%%%%% D1 DD0 term
% 
       + \D_1 \DD_0   \Bigg[
          - ({f^{q}_{1}})^3
\nonumber \\
& \quad
          - 3 {A^{q}_{2}} {f^{q}_{1}}
          - 3 {A^{q}_{1}} {f^{q}_{2}}
          - 6 {\g^{q,1}_{d,1}} {A^{q}_{1}} {f^{q}_{1}}
          - 6 {g^{b,1}_{1}} {A^{q}_{1}} {f^{q}_{1}}
          - \bt_1 {A^{q}_{1}}
          - \bt_0 \Big(
            3 ({f^{q}_{1}})^2
          + 2 {A^{q}_{2}}
          + 8 {\g^{q,1}_{d,1}} {A^{q}_{1}}
\nonumber \\
& \quad
          + 2 {g^{b,1}_{1}} {A^{q}_{1}}
          \Big)
          - 2 \bt_0^2 {f^{q}_{1}}
          + 10 \zeta_3 ({A^{q}_{1}})^3
          + 3 \zeta_2 ({A^{q}_{1}})^2 {f^{q}_{1}}
          + 3 \zeta_2 \bt_0 ({A^{q}_{1}})^2
          \Bigg]
% 
%%%%%%% D1 DD1 term
% 
       + \D_1 \DD_1   \Bigg[
            3 {A^{q}_{1}} ({f^{q}_{1}})^2
\nonumber \\
& \quad
          + 3 {A^{q}_{1}} {A^{q}_{2}}
          + 3 {\g^{q,1}_{d,1}} ({A^{q}_{1}})^2
          + 3 {g^{b,1}_{1}} ({A^{q}_{1}})^2
          + 5 \bt_0 {A^{q}_{1}} {f^{q}_{1}}
          + \bt_0^2 {A^{q}_{1}}
          - \frac{3}{2} \zeta_2 ({A^{q}_{1}})^3
          \Bigg]
% 
%%%%%%%% D2 delta(1-z2) term 
% 
\nonumber \\
& \quad
       + \D_2 \delta(1-z_2)   \Bigg[
          - \frac{1}{2} ({f^{q}_{1}})^3
          - \frac{3}{2} {A^{q}_{2}} {f^{q}_{1}}
          - \frac{3}{2} {A^{q}_{1}} {f^{q}_{2}}
          - 3 {\g^{q,1}_{d,1}} {A^{q}_{1}} {f^{q}_{1}}
          - 3 {g^{b,1}_{1}} {A^{q}_{1}} {f^{q}_{1}}
          - \frac{1}{2} \bt_1 {A^{q}_{1}}
\nonumber \\
& \quad
          - \bt_0 \Big(
            \frac{3}{2} ({f^{q}_{1}})^2
          + {A^{q}_{2}}
          + 4 {\g^{q,1}_{d,1}} {A^{q}_{1}}
          + {g^{b,1}_{1}} {A^{q}_{1}}
          \Big)
          - \bt_0^2 {f^{q}_{1}}
          + 5 \zeta_3 ({A^{q}_{1}})^3
          + \frac{3}{2} \zeta_2 ({A^{q}_{1}})^2 {f^{q}_{1}}
\nonumber \\
& \quad
          + \frac{3}{2} \zeta_2 \bt_0 ({A^{q}_{1}})^2
          \Bigg]
% 
%%%%%%%% D2 DD0 term 
% 
       + \D_2 \DD_0   \Bigg[
            3 {A^{q}_{1}} ({f^{q}_{1}})^2
          + 3 {A^{q}_{1}} {A^{q}_{2}}
          + 3 {\g^{q,1}_{d,1}} ({A^{q}_{1}})^2
          + 3 {g^{b,1}_{1}} ({A^{q}_{1}})^2
          + 5 \bt_0 {A^{q}_{1}} {f^{q}_{1}}
\nonumber \\
& \quad
          + \bt_0^2 {A^{q}_{1}}
          - \frac{3}{2} \zeta_2 ({A^{q}_{1}})^3
          \Bigg]
% 
%%%%%%%% D2 DD1 term 
% 
       + \D_2 \DD_1   \Bigg[
          - \frac{15}{2} ({A^{q}_{1}})^2 {f^{q}_{1}}
          - 5 \bt_0 ({A^{q}_{1}})^2
          \Bigg]
% 
%%%%%%%% D2 DD2 term 
% 
       + \D_2 \DD_2   \Bigg[
            \frac{15}{8} ({A^{q}_{1}})^3
          \Bigg]
\nonumber \\
& \quad
       + \D_3 \delta(1-z_2)   \Bigg[
            {A^{q}_{1}} ({f^{q}_{1}})^2
          + {A^{q}_{1}} {A^{q}_{2}}
          + {\g^{q,1}_{d,1}} ({A^{q}_{1}})^2
          + {g^{b,1}_{1}} ({A^{q}_{1}})^2
          + \frac{5}{3} \bt_0 {A^{q}_{1}} {f^{q}_{1}}
          + \frac{1}{3} \bt_0^2 {A^{q}_{1}}
\nonumber \\
& \quad
          - \frac{1}{2} \zeta_2 ({A^{q}_{1}})^3
          \Bigg]
       + \D_3 \DD_0   \Bigg[
          - \frac{5}{2} ({A^{q}_{1}})^2 {f^{q}_{1}}
          - \frac{5}{3} \bt_0 ({A^{q}_{1}})^2
          \Bigg]
       + \D_3 \DD_1   \Bigg[
            \frac{5}{2} ({A^{q}_{1}})^3
          \Bigg]
\nonumber \\
& \quad
       + \D_4 \delta(1-z_2)    \Bigg[
          - \frac{5}{8} ({A^{q}_{1}})^2 {f^{q}_{1}}
          - \frac{5}{12} \bt_0 ({A^{q}_{1}})^2
          \Bigg]
       + \D_4 \DD_0   \Bigg[
            \frac{5}{8} ({A^{q}_{1}})^3
          \Bigg]
\nonumber \\
& \quad
       + \D_5 \delta(1-z_2)   \Bigg[
           \frac{1}{8} ({A^{q}_{1}})^3
          \Bigg]
       + \Bigg\{ z_1 \leftrightarrow z_2 \Bigg\} \, .
\end{align}
At the stage, we can demonstrate that integration over the rapidity correctly reproduces inclusive threshold contribution to the Higgs production in bottom anti-bottom annihilation reported in \cite{Ahmed:2014cha} :
\begin{eqnarray}
\int dy {d \over dy}\sigma_{b}(\tau,y,q^2)  = \sigma_b(\tau,q^2) \, .
\end{eqnarray}
The integration over the rapidity $y$ leads to the following relation between $\Delta_{d,b}^{\rm SV}(z_1,z_2)$ obtained in this paper and $\Delta^{\rm SV}_{b}(z)$ in \cite{Ahmed:2014cha}:
\begin{eqnarray}
\Delta^{\rm SV}_b(z) = \int dz_1 \int dz_2 \delta(z-z_1 z_2) \Delta^{\rm SV}_{d,b}(z_1,z_2) \, .
\end{eqnarray}
We have explicitly checked that the results presented here for $\Delta^{\rm SV}_{d,b}$ and those for $\Delta^{\rm SV}_{b}$ in the \cite{Ahmed:2014cha} up to N$^3$LO level satisfy the above relation confirming the consistency of the formalism used.  For completeness, we present the results for $\Delta^{{\rm SV},(i)}_{d,b}$ up to N$^3$LO after substituting all the constants that are required to this order:
\begin{align}
&\Delta_{d,b}^{{\rm SV},(1)} =
         \delta(1-z_1) \delta(1-z_2) C_F   (
          - 2
          + 6 \zeta_2
          )
       + \D_0 \DD_0   (
            2 C_F
          )
       + \D_1 \delta(1-z_2)    (
           4 C_F
          )
       + \Bigg\{ z_1 \leftrightarrow z_2 \Bigg\} \, ,
\nonumber \\
%
% %%%%%%%%%%%%%%%%%%%%%%%%%%%%%%%%%%%%%% CI2
%    
&\Delta_{d,b}^{{\rm SV},(2)} =
         \delta(1-z_1) \delta(1-z_2) \Bigg[
         C_F C_A   \Big(
            \frac{83}{9}
          + \frac{32}{3} \zeta_3
          + \frac{250}{9} \zeta_2
          - \frac{26}{5} \zeta_2^2
          \Big)
       +  C_F^2   \Big(
            8
          - 30 \zeta_3
          - 8 \zeta_2
\nonumber \\
& \quad
          + \frac{152}{5} \zeta_2^2
          \Big)
       +  n_f C_F   \Big(
            \frac{4}{9}
          + \frac{4}{3} \zeta_3
          - \frac{40}{9} \zeta_2
          \Big)
       \Bigg]
       + \D_0 \delta(1-z_2) \Bigg[
          C_F C_A   \Big(
          - \frac{808}{27}
          + 28 \zeta_3
          + \frac{44}{3} \zeta_2
          \Big)
\nonumber \\
& \quad
       +  C_F^2   \Big(
            32 \zeta_3
          \Big)
       +  n_f C_F   \Big(
            \frac{112}{27}
          - \frac{8}{3} \zeta_2
          \Big)
       \Bigg]
       + \D_0 \DD_0  \Bigg[
          C_F C_A   \Big(
            \frac{134}{9}
          - 4 \zeta_2
          \Big)
       +  C_F^2   \Big(
          - 8
          + 8 \zeta_2
          \Big)
\nonumber \\
& \quad
       +  n_f C_F   \Big(
          - \frac{20}{9}
          \Big)
       \Bigg]
       + \D_1 \delta(1-z_2) \Bigg[
          C_F C_A   \Big(
            \frac{268}{9}
          - 8 \zeta_2
          \Big)
       + C_F^2   \Big(
          - 16
          + 16 \zeta_2
          \Big)
\nonumber \\
& \quad
       + n_f C_F   \Big(
          - \frac{40}{9}
          \Big)
       \Bigg]
       + \D_1 \DD_0  \Bigg[ 
         C_F C_A   \Big(
          - \frac{44}{3}
          \Big)
       + n_f C_F   \Big(
            \frac{8}{3}
          \Big)
       \Bigg]
       + \D_1 \DD_1  \Bigg[ 
           C_F^2 \Big( 24 \Big)
         \Bigg] 
\nonumber \\
& \quad
       + \D_2 \delta(1-z_2) \Bigg[ 
         C_F C_A   \Big(
          - \frac{22}{3}
          \Big)
       + n_f C_F   \Big(
            \frac{4}{3}
          \Big)
       \Bigg]
       + \D_2 \DD_0 \Bigg[ C_F^2   \Big(
           24
          \Big) \Bigg]
\nonumber \\
& \quad
       + \D_3 \delta(1-z_2) \Bigg[ C_F^2   \Big(
           8
          \Big) \Bigg]
       + \Bigg\{ z_1 \leftrightarrow z_2 \Bigg\} 
\nonumber \\
\intertext{and}
%
% %%%%%%%%%%%%%%%%%%%%%%%%%%%%%%%%%%%%%% CI3
%    
&\Delta_{d,b}^{{\rm SV},(3)} =
        \delta(1-z_1) \delta(1-z_2) \Bigg[
        C_F C_A^2   \Big(
            \frac{34495}{81}
          - 42 \zeta_5
          + \frac{14254}{81} \zeta_3
          - \frac{200}{3} \zeta_3^2
          + \frac{4487}{81} \zeta_2
          - 324 \zeta_2 \zeta_3
\nonumber \\
& \quad
          - \frac{2446}{135} \zeta_2^2
          + \frac{12176}{315} \zeta_2^3
          \Big)
       + C_F^2 C_A   \Big(
          - \frac{491}{3}
          - \frac{2732}{9} \zeta_5
          - \frac{922}{3} \zeta_3
          + \frac{632}{3} \zeta_3^2
          + \frac{10441}{27} \zeta_2
\nonumber \\
& \quad
          + \frac{4288}{9} \zeta_2 \zeta_3
          + \frac{21302}{135} \zeta_2^2
          - \frac{39136}{315} \zeta_2^3
          \Big)
       + C_F^3   \Big(
            \frac{539}{3}
          + 424 \zeta_5
          - 594 \zeta_3
          + \frac{368}{3} \zeta_3^2
          - \frac{179}{3} \zeta_2
\nonumber \\
& \quad
          - 152 \zeta_2 \zeta_3
          - \frac{196}{5} \zeta_2^2
          + \frac{45008}{315} \zeta_2^3
          \Big)
       + n_f C_F C_A   \Big(
          - \frac{5770}{81}
          - 4 \zeta_5
          - \frac{5660}{81} \zeta_3
          - \frac{806}{27} \zeta_2
\nonumber \\
& \quad
          + \frac{136}{3} \zeta_2 \zeta_3
          - \frac{380}{27} \zeta_2^2
          \Big)
       + n_f C_F^2   \Big(
          - \frac{35}{9}
          - \frac{112}{9} \zeta_5
          + 180 \zeta_3
          - \frac{1507}{27} \zeta_2
          - \frac{736}{9} \zeta_2 \zeta_3
          - \frac{1604}{135} \zeta_2^2
          \Big)
\nonumber \\
& \quad
       + n_f^2 C_F   \Big(
            \frac{8}{27}
          - \frac{80}{81} \zeta_3
          + \frac{184}{81} \zeta_2
          + \frac{296}{135} \zeta_2^2
          \Big)
         \Bigg]
% 
%%%%% 
       + \D_0 \delta(1-z_2) \Bigg[
           C_F C_A^2   \Big(
          - \frac{297029}{729}
          - 192 \zeta_5
\nonumber \\
& \quad
          + \frac{14264}{27} \zeta_3
          + \frac{27752}{81} \zeta_2
          - \frac{176}{3} \zeta_2 \zeta_3
          - \frac{616}{15} \zeta_2^2
          \Big)
       + C_F^2 C_A   \Big(
            \frac{3232}{27}
          + \frac{3280}{9} \zeta_3
          - \frac{4816}{27} \zeta_2
          - 16 \zeta_2 \zeta_3
\nonumber \\
& \quad
          + \frac{176}{3} \zeta_2^2
          \Big)
       + C_F^3   \Big(
            384 \zeta_5
          - 128 \zeta_3
          - 128 \zeta_2 \zeta_3
          \Big)
       + n_f C_F C_A   \Big(
            \frac{62626}{729}
          - \frac{536}{9} \zeta_3
          - \frac{7760}{81} \zeta_2
\nonumber \\
& \quad
          + \frac{208}{15} \zeta_2^2
          \Big)
       + n_f C_F^2   \Big(
            \frac{421}{9}
          - \frac{944}{9} \zeta_3
          + \frac{520}{27} \zeta_2
          - \frac{256}{15} \zeta_2^2
          \Big)
       + n_f^2 C_F   \Big(
          - \frac{1856}{729}
          - \frac{32}{27} \zeta_3
\nonumber \\
& \quad
          + \frac{160}{27} \zeta_2
          \Big)
         \Bigg]
% 
%%%%%
       + \D_0 \DD_0 \Bigg[
        C_F C_A^2   \Big(
            \frac{15503}{81}
          - 88 \zeta_3
          - \frac{340}{3} \zeta_2
          + \frac{88}{5} \zeta_2^2
          \Big)
       + C_F^2 C_A   \Big(
          - \frac{68}{3}
          - \frac{400}{3} \zeta_3
\nonumber \\
& \quad
          + \frac{608}{9} \zeta_2
          - \frac{24}{5} \zeta_2^2
          \Big)
       + C_F^3   \Big(
            32
          - 120 \zeta_3
          + 32 \zeta_2
          - \frac{96}{5} \zeta_2^2
          \Big)
       + n_f C_F C_A   \Big(
          - \frac{4102}{81}
          + \frac{256}{9} \zeta_2
          \Big)
\nonumber \\
& \quad
       + n_f C_F^2   \Big(
          - \frac{23}{3}
          + \frac{160}{3} \zeta_3
          - \frac{80}{9} \zeta_2
          \Big)
       + n_f^2 C_F   \Big(
            \frac{200}{81}
          - \frac{16}{9} \zeta_2
          \Big)
          \Bigg] 
% 
%%%%%%
       + \D_1 \delta(1-z_2) \Bigg[
         C_F C_A^2   \Big(
            \frac{31006}{81}
\nonumber \\
& \quad
          - 176 \zeta_3
          - \frac{680}{3} \zeta_2
          + \frac{176}{5} \zeta_2^2
          \Big)
       + C_F^2 C_A   \Big(
          - \frac{136}{3}
          - \frac{800}{3} \zeta_3
          + \frac{1216}{9} \zeta_2
          - \frac{48}{5} \zeta_2^2
          \Big)
       + C_F^3   \Big(
            64
\nonumber \\
& \quad
          - 240 \zeta_3
          + 64 \zeta_2
          - \frac{192}{5} \zeta_2^2
          \Big)
       + n_f C_F C_A   \Big(
          - \frac{8204}{81}
          + \frac{512}{9} \zeta_2
          \Big)
       + n_f C_F^2   \Big(
          - \frac{46}{3}
          + \frac{320}{3} \zeta_3
\nonumber \\
& \quad
          - \frac{160}{9} \zeta_2
          \Big)
       + n_f^2 C_F   \Big(
           \frac{400}{81}
          - \frac{32}{9} \zeta_2
          \Big)
         \Bigg]
% 
%%%%%%%%%
       + \D_1 \DD_0  \Bigg[
        C_F C_A^2   \Big(
          - \frac{7120}{27}
          + \frac{176}{3} \zeta_2
          \Big)
\nonumber \\
& \quad
       + C_F^2 C_A   \Big(
          - \frac{2704}{9}
          + 336 \zeta_3
          + 352 \zeta_2
          \Big)
       + C_F^3   \Big(
            640 \zeta_3
          \Big)
       + n_f C_F C_A   \Big(
            \frac{2312}{27}
          - \frac{32}{3} \zeta_2
          \Big)
\nonumber \\
& \quad
       + n_f C_F^2   \Big(
            \frac{424}{9}
          - 64 \zeta_2
          \Big)
       + n_f^2 C_F   \Big(
          - \frac{160}{27}
          \Big)
         \Bigg]
% 
%%%%%%%%
       + \D_1 \DD_1  \Bigg[
         C_F C_A^2   \Big(
            \frac{484}{9}
          \Big)
       + C_F^2 C_A   \Big(
            \frac{1072}{3}
\nonumber \\
& \quad
          - 96 \zeta_2
          \Big)
       + C_F^3   \Big(
          - 96
          - 96 \zeta_2
          \Big)
       + n_f C_F C_A   \Big(
          - \frac{176}{9}
          \Big)
       + n_f C_F^2   \Big(
          - \frac{160}{3}
          \Big)
       + n_f^2 C_F   \Big(
            \frac{16}{9}
          \Big)
         \Bigg]
\nonumber \\
& \quad
%%%%%%%%
       + \D_2 \delta(1-z_2)  \Bigg[
         C_F C_A^2   \Big(
          - \frac{3560}{27}
          + \frac{88}{3} \zeta_2
          \Big)
       + C_F^2 C_A   \Big(
          - \frac{1352}{9}
          + 168 \zeta_3
          + 176 \zeta_2
          \Big)
\nonumber \\
& \quad
       + C_F^3   \Big(
            320 \zeta_3
          \Big)
       + n_f C_F C_A   \Big(
            \frac{1156}{27}
          - \frac{16}{3} \zeta_2
          \Big)
       + n_f C_F^2   \Big(
            \frac{212}{9}
          - 32 \zeta_2
          \Big)
       + n_f^2 C_F   \Big(
          - \frac{80}{27}
          \Big)
         \Bigg]
\nonumber \\
& \quad
%%%%%%
       + \D_2 \DD_0  \Bigg[
         C_F C_A^2   \Big(
            \frac{484}{9}
          \Big)
       + C_F^2 C_A   \Big(
            \frac{1072}{3}
          - 96 \zeta_2
          \Big)
       + C_F^3   \Big(
          - 96
          - 96 \zeta_2
          \Big)
\nonumber \\
& \quad
       + n_f C_F C_A   \Big(
          - \frac{176}{9}
          \Big)
       + n_f C_F^2   \Big(
          - \frac{160}{3}
          \Big)
       + n_f^2 C_F   \Big(
            \frac{16}{9}
          \Big)
         \Bigg]
% 
%%%%%%%%
       + \D_2 \DD_1   \Bigg[
         C_F^2 C_A   \Big(
          - \frac{880}{3}
          \Big)
\nonumber \\
& \quad
       + n_f C_F^2   \Big(
            \frac{160}{3}
          \Big)
         \Bigg]
% 
%%%%%%%
       + \D_2 \DD_2  \Bigg[
         C_F^3   \Big(
            120
          \Big)
          \Bigg]
% 
%%%%%%
       + \D_3 \delta(1-z_2)  \Bigg[
         C_F C_A^2   \Big(
            \frac{484}{27}
          \Big)
       + C_F^2 C_A   \Big(
            \frac{1072}{9}
\nonumber \\
& \quad
          - 32 \zeta_2
          \Big)
       + C_F^3   \Big(
          - 32
          - 32 \zeta_2
          \Big)
       + n_f C_F C_A   \Big(
          - \frac{176}{27}
          \Big)
       + n_f C_F^2   \Big(
          - \frac{160}{9}
          \Big)
       + n_f^2 C_F   \Big(
            \frac{16}{27}
          \Big)
         \Bigg]
\nonumber \\
& \quad
%%%%%%
       + \D_3 \DD_0  \Bigg[
         C_F^2 C_A   \Big(
          - \frac{880}{9}
          \Big)
       + n_f C_F^2   \Big(
            \frac{160}{9}
          \Big)
       \Bigg]
% 
%%%%%%
       + \D_3 \DD_1  \Bigg[
         C_F^3   \Big(
            160
          \Big)
       \Bigg]
\nonumber \\
& \quad
%%%%%%%
       + \D_4 \delta(1-z_2)  \Bigg[
         C_F^2 C_A   \Big(
          - \frac{220}{9}
          \Big)
       + n_f C_F^2   \Big(
            \frac{40}{9}
          \Big)
       \Bigg]
       + \D_4 \DD_0  \Bigg[
         C_F^3   \Big(
            40
          \Big)
       \Bigg]
\nonumber \\
& \quad
%%%%%%%
       + \D_5 \delta(1-z_2)  \Bigg[
         C_F^3   \Big(
            8
          \Big)
       \Bigg]
       + \Bigg\{ z_1 \leftrightarrow z_2 \Bigg\} \, .
\end{align}
Substituting $\Delta_{d,b}^{{\rm SV},(1)}$, $\Delta_{d,b}^{{\rm SV},(2)}$ and $\Delta_{d,b}^{{\rm SV},(3)}$ in the Eq.~\ref{Wb}, we obtain $W^{{\rm SV},(i)}_{b}$ or equivalently $\frac{d}{dy}\sigma^{b, {\rm SV}, (i)}$ (Eq.~\ref{sigmaW}) at the hadronic level order by order up to ${\cal O}(a_{s}^{3})$.
\subsection{Numerical Results}
\label{NumRes}
In this section, we present the numerical impact of the rapidity distribution of the Higgs boson, produced via bottom anti-bottom annihilation subprocess at the LHC. The rapidity distribution can be expanded in powers of the strong coupling constant $a_s$ as 
\begin{align}
{d \sigma^b\over  dY} &={d \sigma^{b,(0)} \over dY} + \sum_{i=1}^\infty a_s^i 
~{d \sigma^{b,(i)} \over dY} \,.
\end{align}
%
% *********** Fig.1 **********
%
\begin{figure}[h!]
\centerline{ 
\includegraphics[width=8cm]{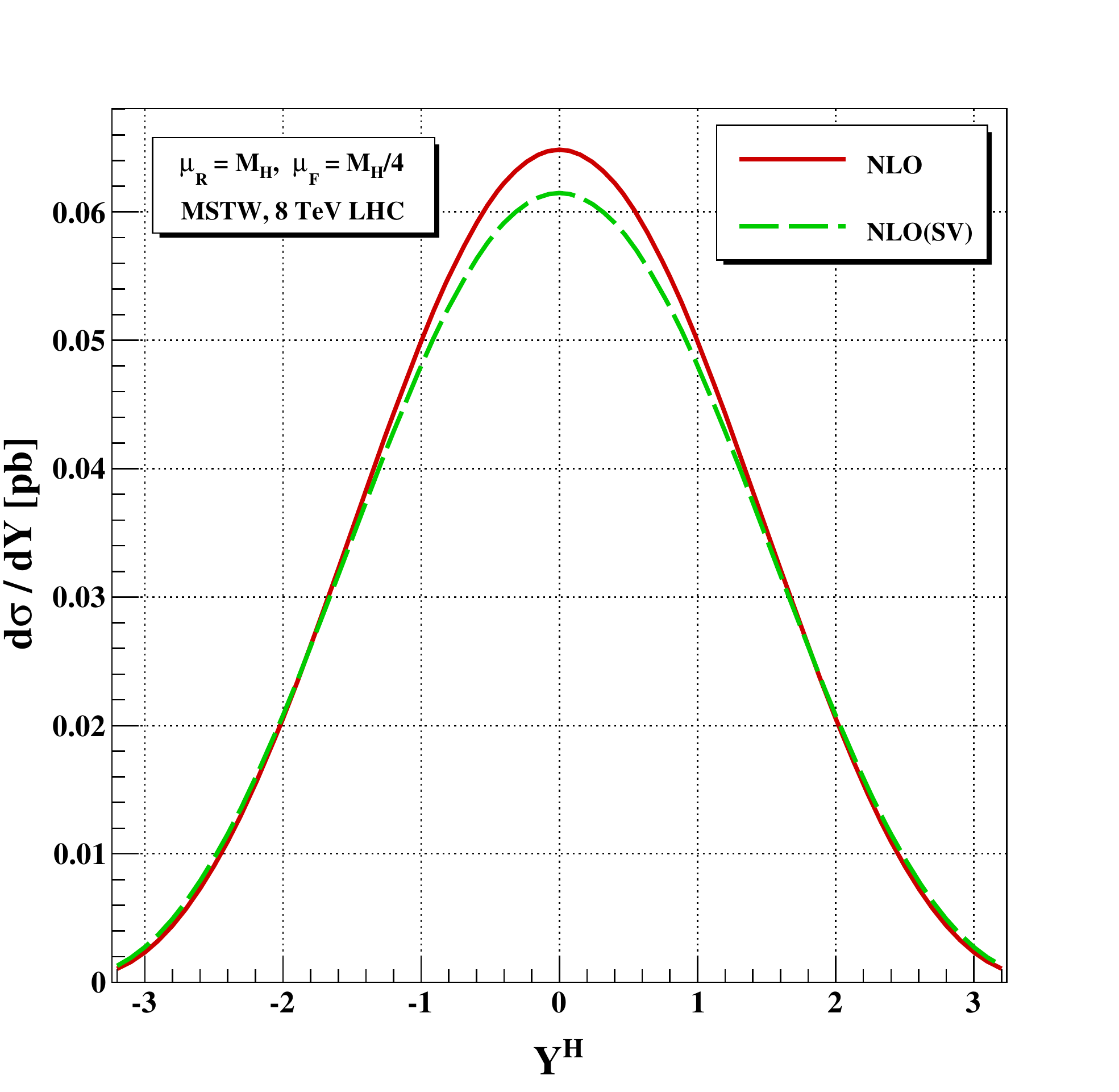}
\includegraphics[width=8cm]{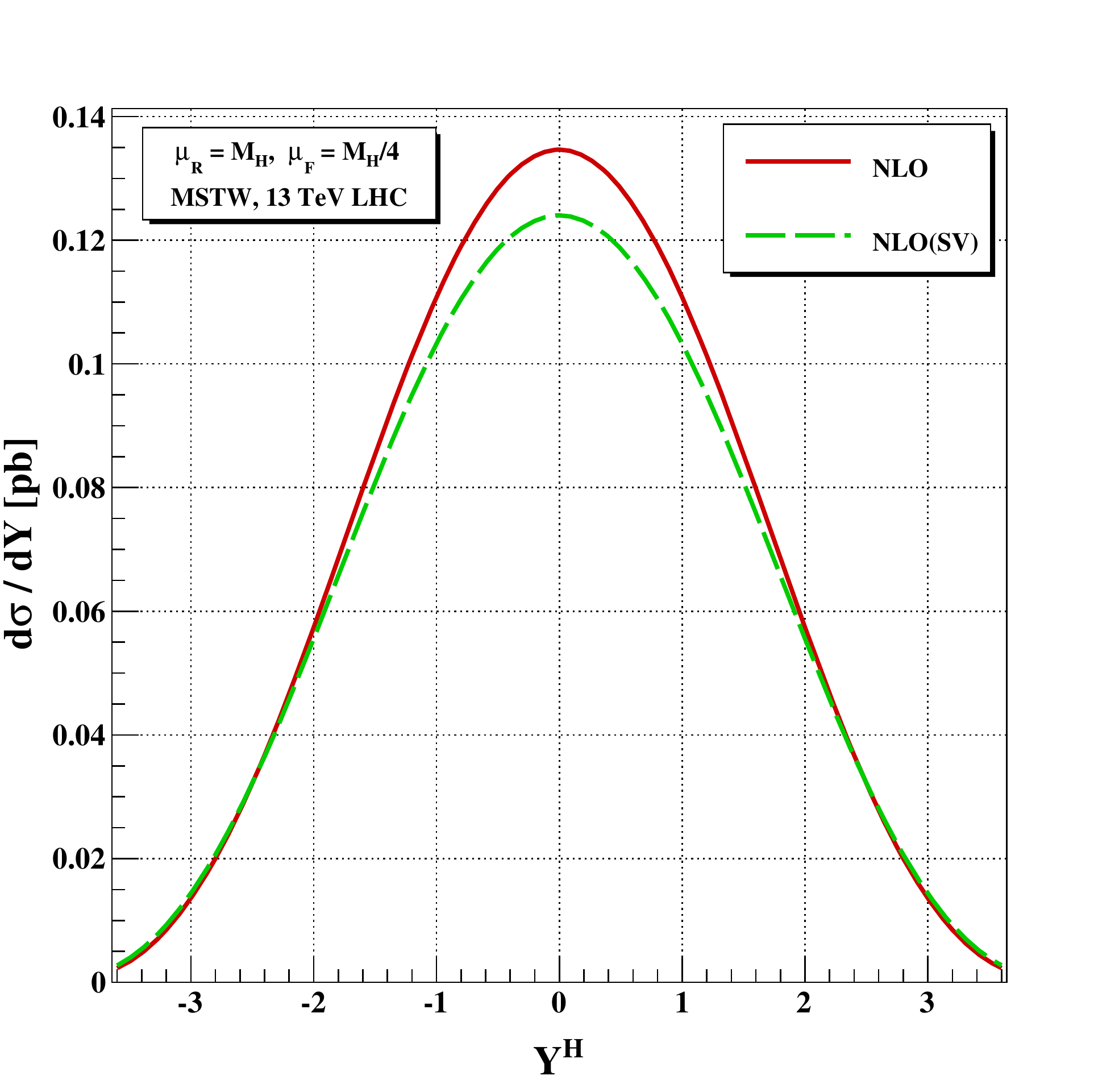}
}
\caption{\label{Fig:NLOSV} The comparison between NLO and NLO$_{\rm SV}$ with the renormalization scale $\mu_{R}=m_{H}$ and factorization scale  $\mu_{F}=m_{H}/4$ at 8 TeV(left panel) and 13 TeV (right panel) LHC.
}
\end{figure}
Beyond LO, the distribution is split into hard and SV parts as
\begin{align}
{d \sigma^{b,(i)} \over dY} = {d \sigma^{{\rm hard},b,(i)} \over dY}
+{d \sigma^{{\rm SV},b,(i)} \over dY}\,.
\end{align}
In the following, for our numerical study we will use the exact results up to NLO level but at NNLO, we use exact NLO and only threshold contribution at ${\cal O}(a_s^2)$ as we do not have access to the hard part at ${\cal O}(a_s^2)$ computed in \cite{Buehler:2012cu} \footnote{The authors informed us that the code is not yet ready for public distribution}.  We call it NNLO(SV).  Similarly at N$^3$LO level, we will use NNLO(SV) and threshold contribution at ${\cal O}(a_s^3)$, denoted by N$^3$LO(SV) hereafter. We present results for the center of mass energies 8 and 13 TeV at the LHC. The standard model parameters which enter into our computation are the Z boson mass $m_{Z} = 91.1876$ GeV, top quark mass $m_{t} = 173.4$ GeV and mass of the Higgs boson $m_{H} = 125$ GeV. The strong coupling constant is evolved using the 4-loop RG equations with $\alpha_s^{\text{N$^3$LO}} (m_Z ) = 0.117$. Following the Ref.~\cite{Vermaseren:1997fq}, the solution to RGE \ref{RGElambda} for $\lambda(\mu_R^2)$ is given by,
\begin{eqnarray}
\lambda(\mu_R^2) = \lambda(\mu_{0}^2) \frac{M(a_s(\mu_R^2))}{M(a_s(\mu_{0}^2))}
\end{eqnarray}
with
\begin{eqnarray}
M(a_s) &=& a_s^{A_0} \sum_{i=0}^\infty a_s^i~M_i\,.
\end{eqnarray}
The $K_i$ are given by 
\begin{eqnarray}
M_0&=&1,\quad \quad M_1=A_1,\quad \quad
\nonumber\\[2ex]
M_2&=&{1 \over 2} (A_1^2+A_2),
\quad \quad M_3={1 \over 6} (A_1^3+3 A_1 A_2+2 A_3),
\end{eqnarray}
with
\begin{eqnarray}
      A_0&=&c_0, \quad \quad
      A_1=c_1-b_1 c_0, \quad \quad
      A_2=c_2-b_1 c_1+c_0 (b_1^2-b_2),
\nonumber\\[2ex]
      A_3&=&c_3-b_1 c_2+c_1 (b_1^2-b_2)+c_0 (b_1 b_2-b_1 (b_1^2-b_2)-b_3),
\end{eqnarray}
and 
\begin{eqnarray}
c_i={\gamma_i^{b} \over \beta_0},\quad \quad  b_i={\gamma_i^{b} \over \beta_0},
\end{eqnarray}
\begin{table}[t]
\begin{center}
\begin{tabular}{ l  c  c  c  c  c  c  c  c  c }
    \hline \hline
    $Y$ & ~0.0~ & ~0.4~ & ~0.8~ & ~1.2~ & ~1.6~ & ~2.0~ & ~2.4~ & ~2.8~ & ~3.2~  \\
    \hline 
    $10^{2}$~LO & 4.137 & 4.027 & 3.705 & 3.196 & 2.549 & 1.828 & 1.126 & 5.427 & 1.686   \\
    \hline
    $10^{2}$~NLO & 6.485 & 6.225 & 5.495 & 4.429 & 3.217 & 2.054 & 1.097 & 4.419 & 1.065  \\
    \hline
    $10^{2}$~NNLO(SV) & 6.921 & 6.650 & 5.879 & 4.731 & 3.407 & 2.135 & 1.113 & 4.417 & 1.118  \\
    \hline
    $10^{2}$~N$^3$LO(SV)  & 6.984 & 6.707 & 5.922 & 4.757 & 3.415 & 2.130 & 1.105 & 4.340 & 1.084   \\
    \hline \hline
 \end{tabular}
 \caption{Contributions at LO, NLO, NNLO(SV) and N$^3$LO(SV) with the renormalization scale $\mu_{R}=m_{H}$ and factorization scale  $\mu_{F}=m_{H}/4$ at 8 TeV LHC.}
 \label{table:central_mu_LHC8}
 \end{center}
\end{table}
\begin{table}[t]
\begin{center}
\begin{tabular}{ l  c  c  c  c  c  c  c  c  c }
    \hline \hline
    $Y$ & ~0.0~ & ~0.4~ & ~0.8~ & ~1.2~ & ~1.6~ & ~2.0~ & ~2.4~ & ~2.8~ & ~3.2~  \\
    \hline 
    $10^{2}$~LO & 8.465 & 8.293 & 7.787 & 6.981 & 5.925 & 4.686 & 3.371 & 2.115 & 1.068   \\
    \hline
    $10^{2}$~NLO & 13.466 & 13.063 & 11.903 & 10.133 & 7.985 & 5.737 & 3.671 & 2.001 & 0.849 \\
    \hline
    $10^{2}$~NNLO(SV) & 14.284 & 13.875 & 12.689 & 10.844 & 8.549 & 6.099 & 3.833 & 2.035 & 0.848  \\
    \hline
    $10^{2}$~N$^3$LO(SV)  & 14.475 & 14.057 & 12.843 & 10.959 & 8.620 & 6.131 & 3.837 & 2.025 & 0.838  \\
    \hline \hline
 \end{tabular}
 \caption{Contributions at LO, NLO, NNLO(SV) and N$^3$LO(SV) with the renormalization scale $\mu_{R}=m_{H}$ and factorization scale  $\mu_{F}=m_{H}/4$ at 13 TeV LHC.}
 \label{table:central_mu_LHC13}
 \end{center}
\end{table}
where $\mu_{0}$ is some reference scale at which $\lambda$ is known. We have numerically evaluated $\lambda(\mu_R^2)$ to relevant order namely LO, NLO, NNLO and N$^{3}$LO by truncating the terms in the RHS of Eq.~\ref{RGElambda}. We have used  $\lambda(\mu^{2}_{0})= \sqrt 2m_b (\mu_{0})/v$ and $m_b (\mu_{0}) = 3.63$ GeV with the choice $\mu_{0} = 10$ GeV. We use the MSTW2008~\cite{Martin:2009iq} parton density sets with errors estimated at 68$\%$ confidence level with five active flavours. Parton densities and $\alpha_{s}$ are evaluated at each corresponding perturbative order. Specifically, we use $(n + 1)$-loop $\alpha_{s}$ at N$^{n}$LO, with $n = 0, 1, 2, 3$. However, we use MSTW2008NNLO PDFs at N$^{3}$LO, the N$^{3}$LO kernels not being available at the moment. We set the renormalization scale $\mu_R = m_H$ and factorization scale $\mu_F = m_{H}/4$ \cite{Maltoni:2003pn} as their central values. 

Several checks have been performed on our numerical code. We have found complete agreement with the literature on the inclusive Higgs production rate \cite{Harlander:2003ai, Ahmed:2014cha} after performing an additional numerical integration over the rapidity Y of our distribution.  The check was also performed at the analytical level. However, we were not able to reproduce the plot given in \cite{Buehler:2012cu}, after using the same set of values of the input parameters.
%
% *********** Fig.2 **********
%
\begin{figure}[t]
\centerline{ 
\includegraphics[width=8cm]{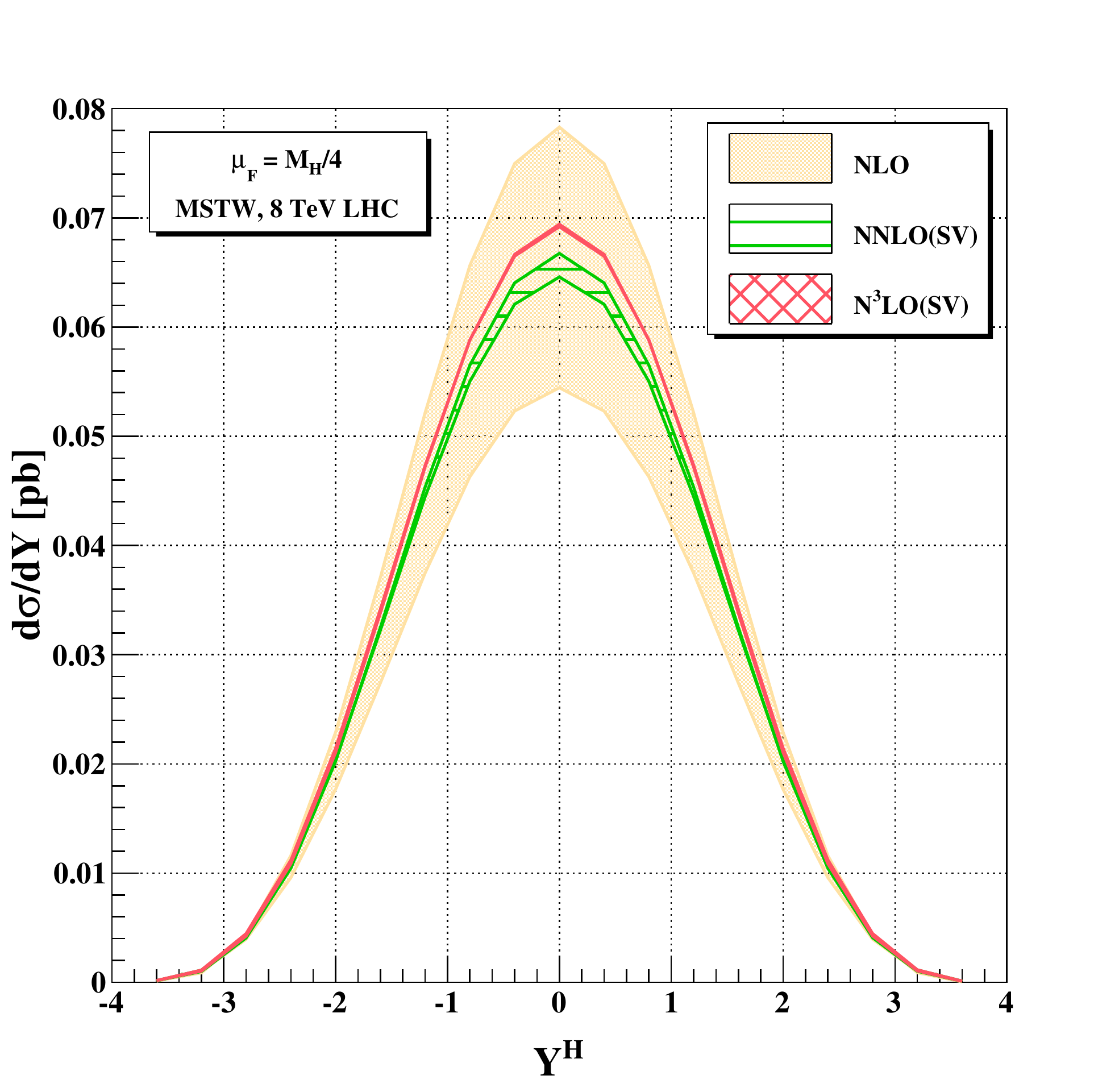}
\includegraphics[width=8cm]{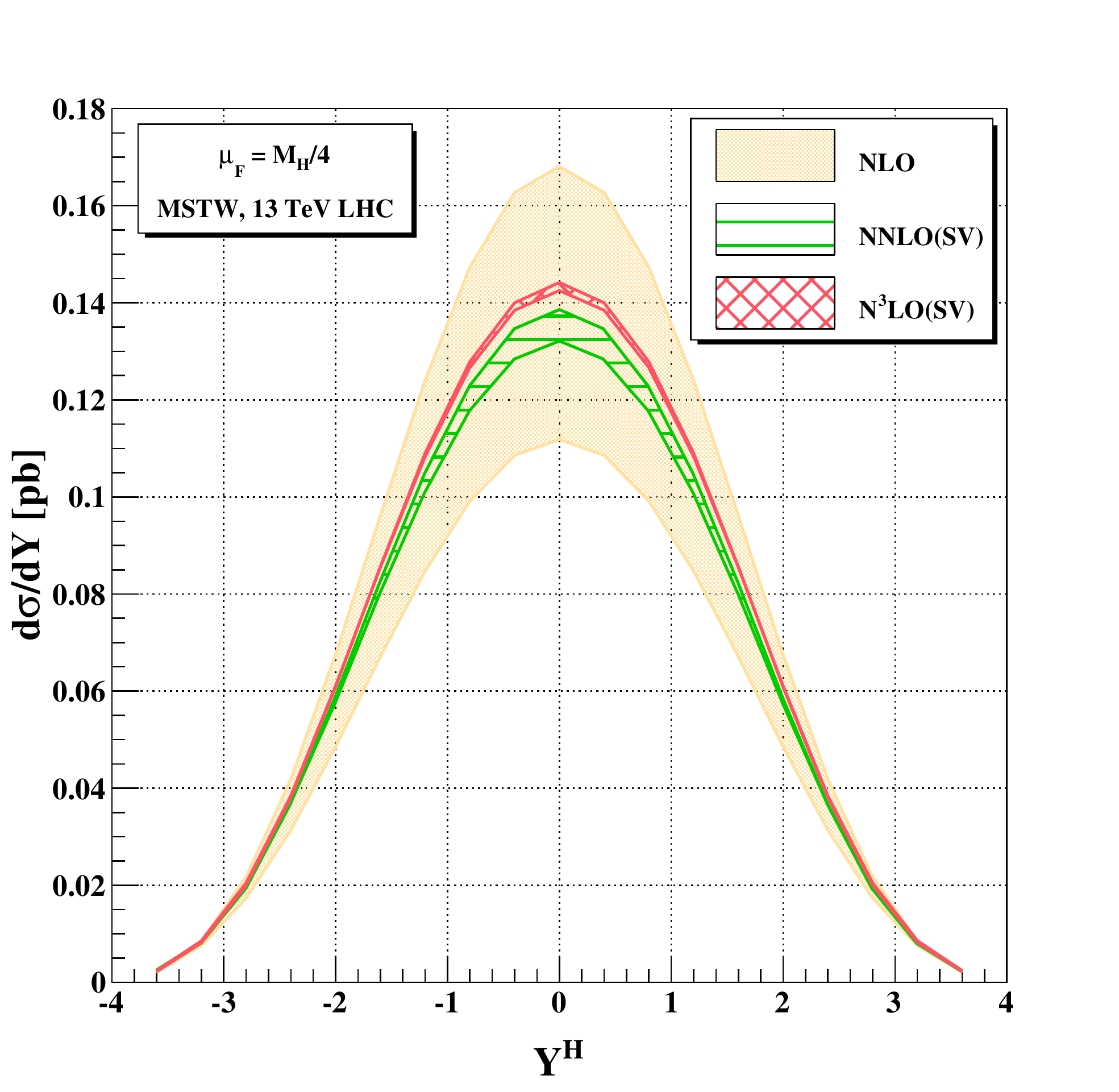}
}
\caption{\label{Fig:murBandplot} The rapidity distribution of the Higgs boson at NLO, NNLO(SV) and N$^{3}$LO(SV) at 8 TeV(left panel) and 13 TeV (right panel) LHC. The band indicates the uncertainty due to renormalization scale.}
\end{figure}
%
% *********** Fig.3 **********
%
\begin{figure}[h]
\centerline{ 
\includegraphics[width=8cm]{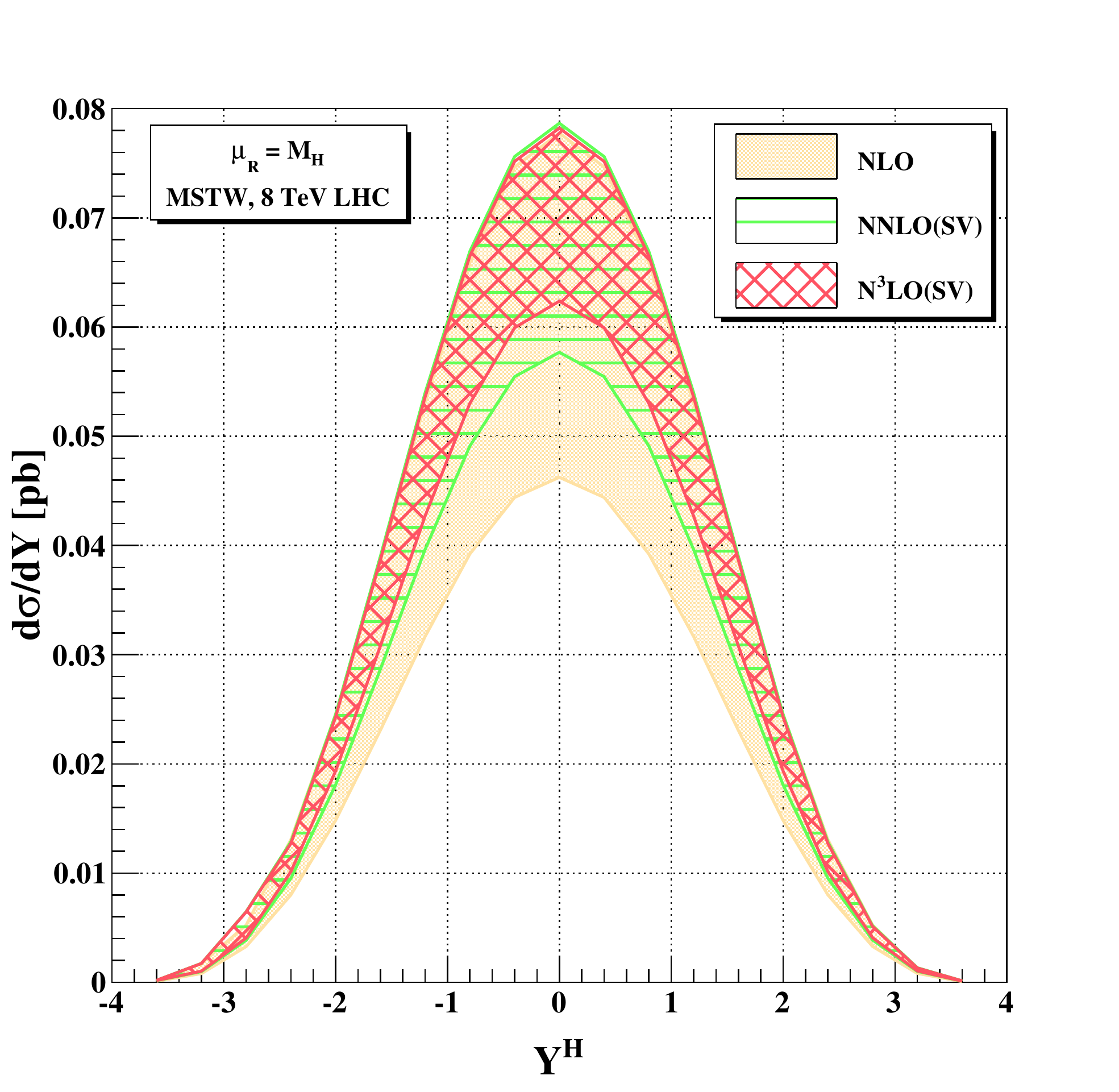}
\includegraphics[width=8cm]{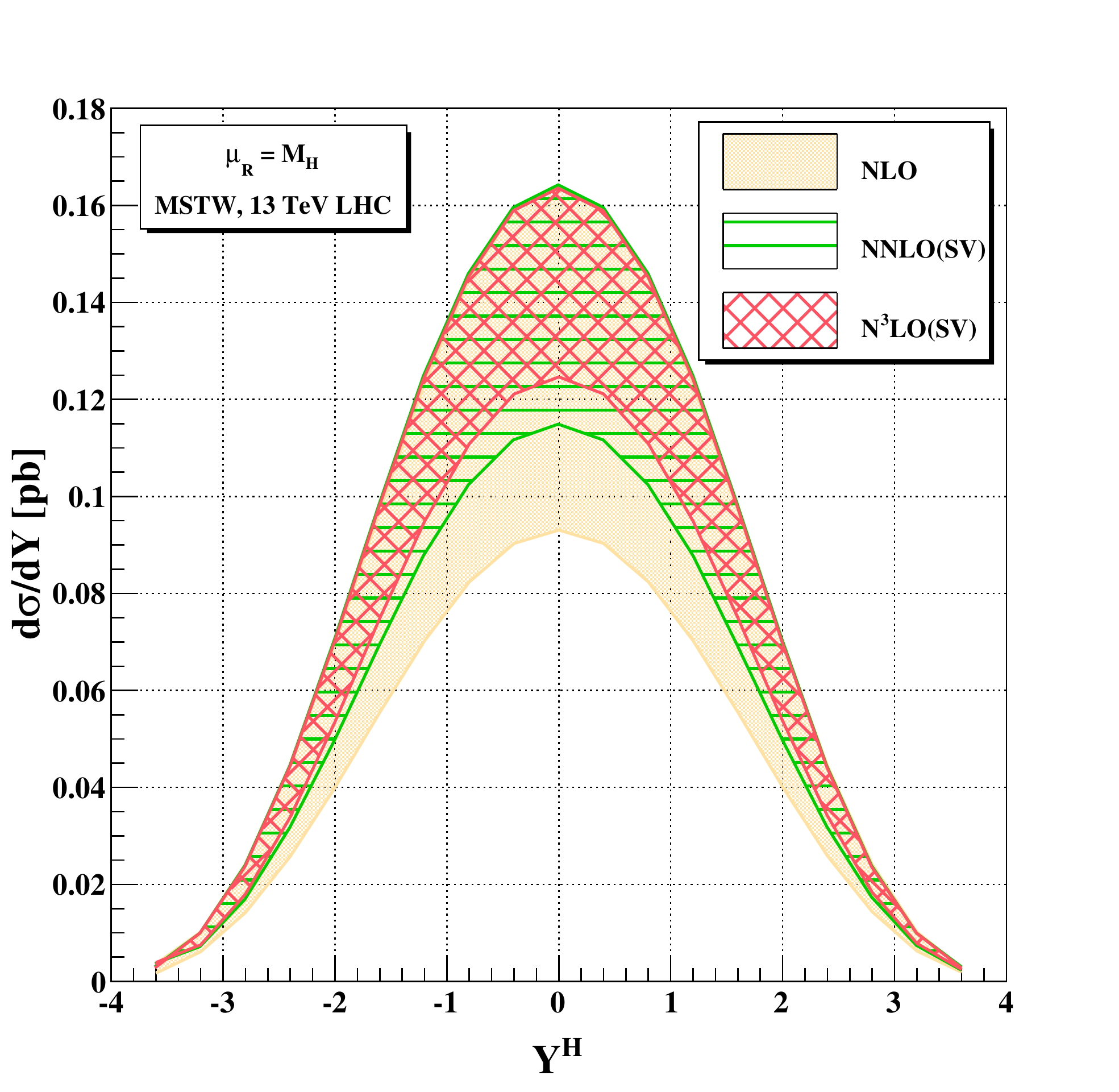}
}
\caption{\label{Fig:mufBandplot} The rapidity distribution of the Higgs boson at NLO, NNLO(SV) and N$^{3}$LO(SV) at 8 TeV(left panel) and 13 TeV (right panel) LHC. The band indicates the uncertainty due to factorization scale.}
\end{figure}
We begin our discussion with the results at NLO level.  In Sec.~\ref{ThresNLO}, we presented the contributions coming from the exact results, containing the regular as well as pure threshold ones to the rapidity distribution at ${\cal O }({a_s})$. In Fig.~\ref{Fig:NLOSV}, we plot both the NLO(SV) and exact NLO rapidity distributions to exhibit the dominance of threshold  over the entire rapidity range after setting the values of the renormalization and factorization scales to their central values. From now onward, we adopt a consistent representation to display the figures corresponding to our results. In every figure, the left panel shows the result for 8 TeV whereas the right panel corresponds to 13 TeV at the LHC.
%
% *********** Fig.4 **********
%
\begin{figure}[t]
\centerline{ 
\includegraphics[width=8cm]{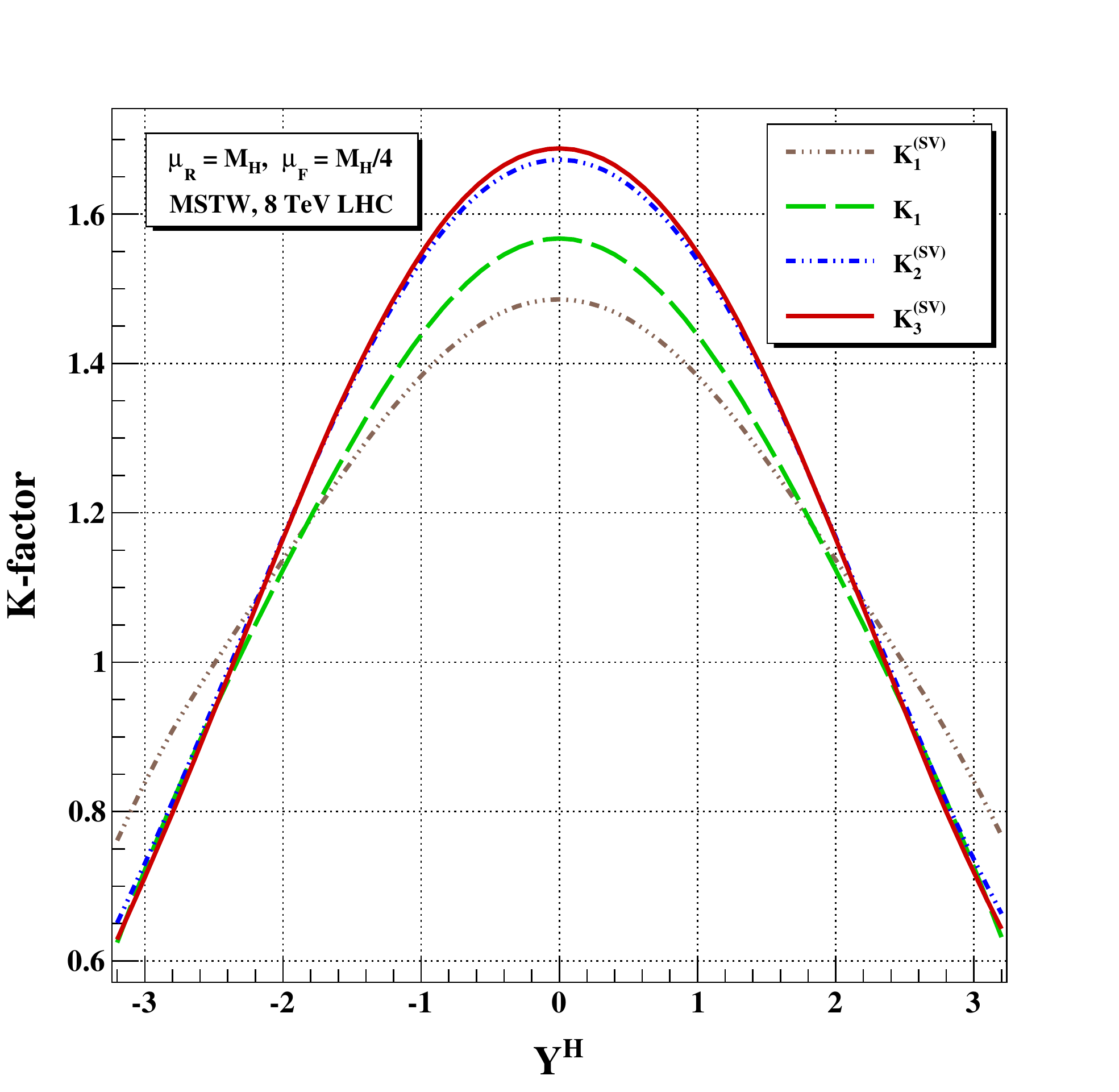}
\includegraphics[width=8cm]{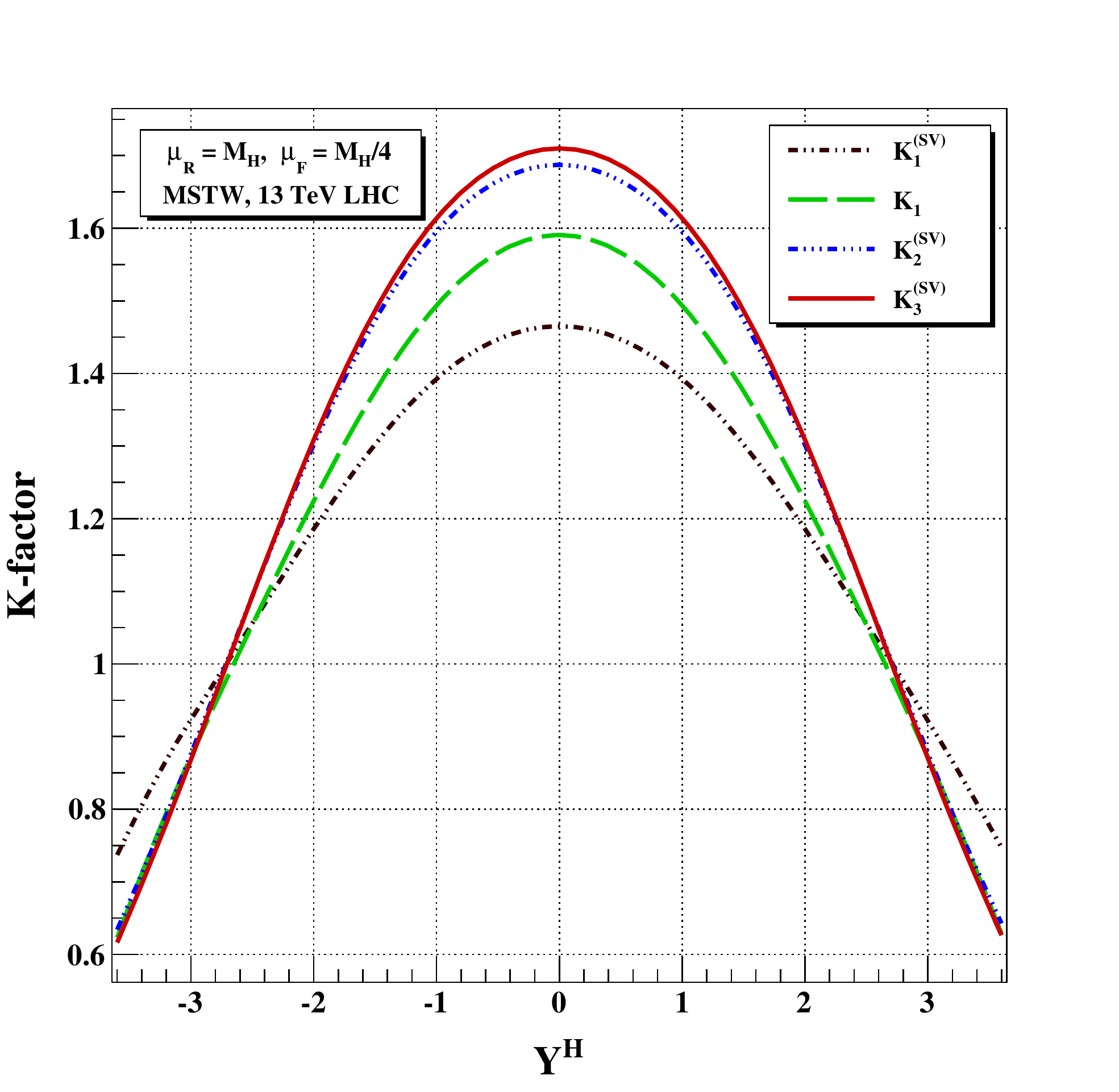}
}
\caption{\label{Fig:Kfactor} The distribution of $K_{1}^{({\rm SV})}$, $K_{1}$, $K_{2}^{({\rm SV})}$ and $K_{3}^{({\rm SV})}$ at different perturbative order at 8 TeV(left panel) and 13 TeV (right panel) LHC.}
\end{figure}
We observe that the exact NLO contribution is well approximated by the NLO(SV), thanks to the intrinsic property of the matrix element, where the phase-space points corresponding to the born kinematics contribute towards the largest radiative corrections for the low  $\tau~(m^{2}_{H}/s \approx 10^{-4})$ values. So, we expect that the trend of approximating the exact results by threshold corrections at that order to remain same after the inclusion of higher order terms also.

With this in mind, we present the results at LO, NLO, NNLO(SV), N$^{3}$LO(SV) for different values of the rapidity Y after setting the central values for renormalization and factorization scales for 8 TeV in Table~\ref{table:central_mu_LHC8} and for 13 TeV in Table~\ref{table:central_mu_LHC13} at LHC. The hadronic cross-section, obtained by the convolution of the partonic cross section with the parton densities, suffers from the theoretical  uncertainties, arising from the missing higher order corrections, through the renormalization ($\mu_R$) and factorization ($\mu_F$) scales. These can be estimated through the variation of the differential hadronic cross section with $\mu_R$ and $\mu_F$, thereby exhibiting the  size of the higher order effects. 

In Fig.~\ref{Fig:murBandplot}, we plot two curves for each order for the predictions at NLO, NNLO({\rm SV}), N$^{3}$LO(SV) corresponding to two different choices of the renormalization scale, $\mu_R = 0.1m_{H}$ and $\mu_R = 10m_{H}$, keeping the factorization scale fixed at $\mu_F = m_{H}/4$, whereas in Fig.~\ref{Fig:mufBandplot}, we plot the predictions at each order corresponding to two different choices of the factorization scale, $\mu_{F} = m_{H}/8$ and $\mu_{F} = m_{H}/2$, keeping the renormalization scale fixed at $\mu_{R}=m_{H}$. We observe a consistent improvement in the accuracy of the predictions with the inclusion of the higher order terms, the width of the bands being an clear indicator of the theoretical uncertainties. Moreover, we can see that the dependence on the renormalization scale for this process is very mild. Another way to assess the reliability of the prediction is to study the rate of convergence of the perturbation series, represented by the K-factor. 

\begin{figure}[t]
\centerline{ 
\includegraphics[width=8cm]{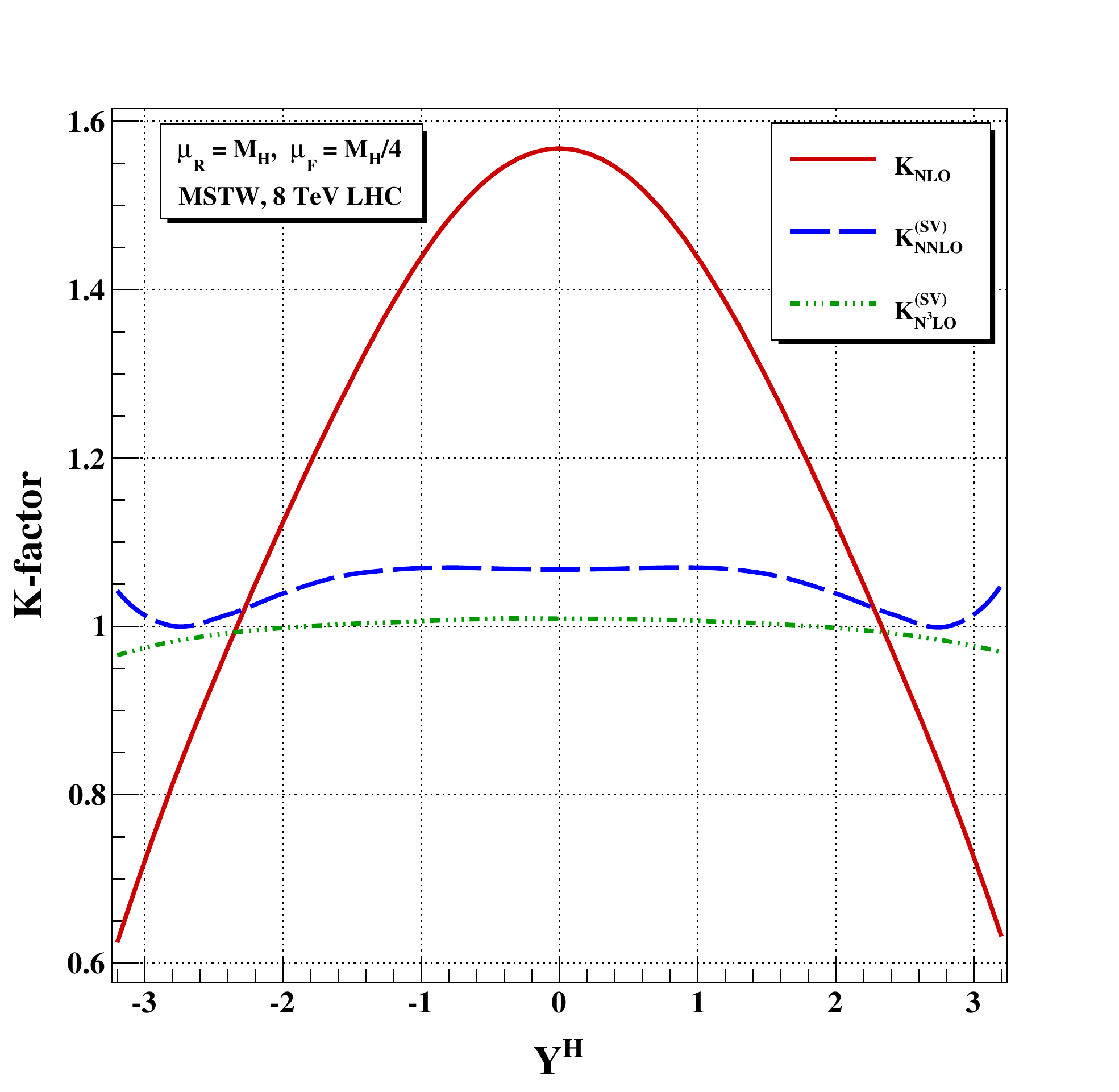}
\includegraphics[width=8cm]{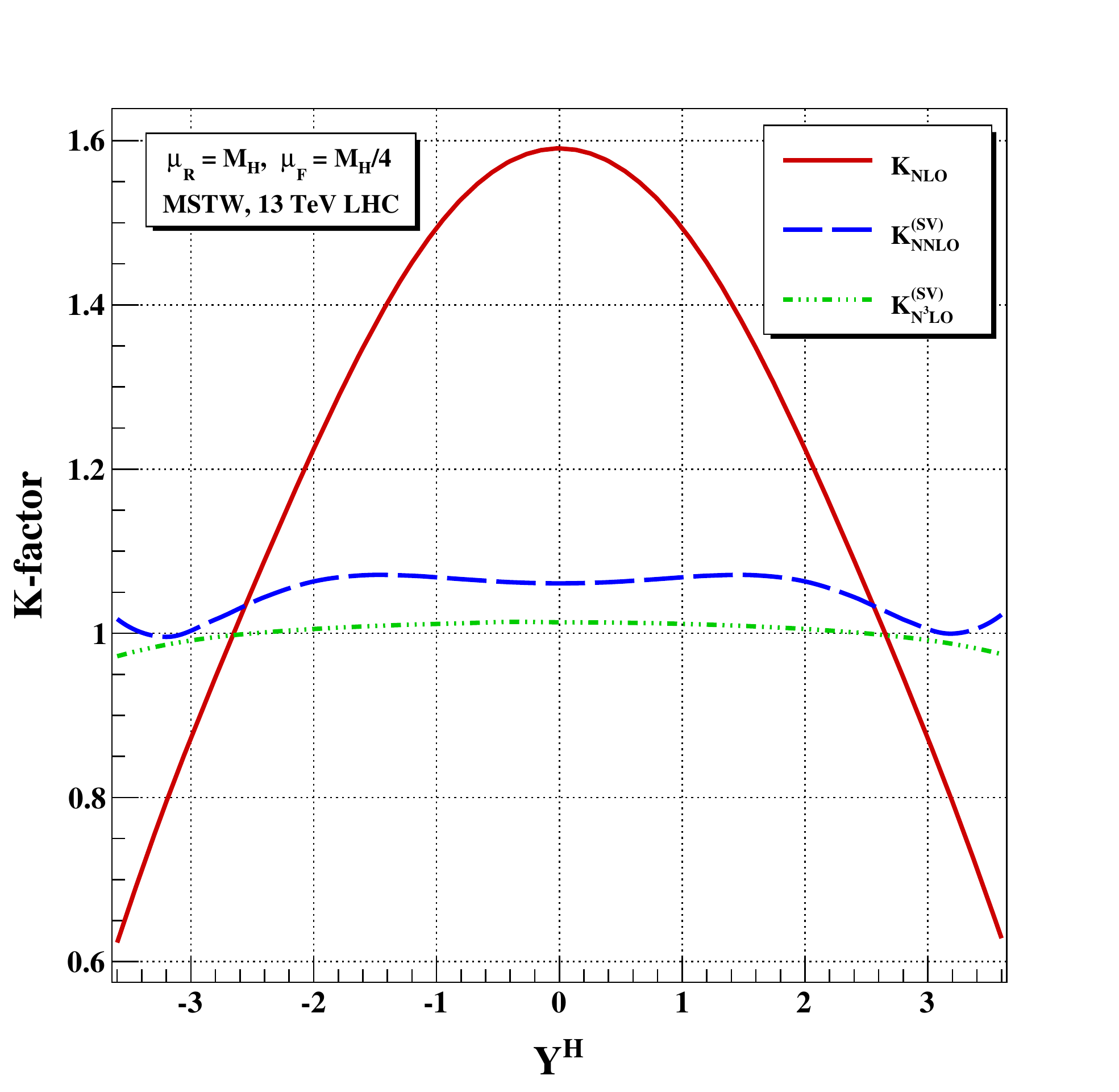}
}
\caption{\label{Fig:KfactorRel} The distribution of $K_{\rm NLO}$, $K_{\rm NNLO}^{({\rm SV})}$ and $K_{\rm N^{3}LO}^{(\rm SV)}$ at different perturbative order at 8 TeV(left panel) and 13 TeV (right panel) LHC.
}
\end{figure}

In the Fig.~\ref{Fig:Kfactor}, we plot the K-factors defined as $K_1={d\sigma^{NLO}}/{d\sigma^{LO}}$ and $ K_{i}^{({\rm SV})} = {d\sigma^{N^iLO(SV)}}/{d\sigma^{LO}}, i=2,3 $ as a function of $Y$. For 8 TeV LHC, we see that the $K_1$ varies from 1.57 to 0.63 over the entire rapidity range, while the value of $K_1$ for the inclusive rate is 1.37. Similarly, for $K_2^{({\rm SV})}$,the variation is from 1.67 to 0.66, while for the inclusive rate it is 1.35. It shows, particularly, that the shape at higher orders can not be rescaled from lower orders as the differential  K-factor varies significantly over the full rapidity range. In the Fig.~\ref{Fig:KfactorRel} we plot K factors defined by $K_{NLO}^{({\rm SV})}=d\sigma^{NLO(SV)}/d\sigma^{LO}, K_{NNLO}^{({\rm SV})}=d\sigma^{NNLO(SV)}/d\sigma^{NLO}$ and $K_{N^3LO}^{({\rm SV})}=d\sigma^{N^3LO(SV)}/d\sigma^{NNLO(SV)}$. The values of the K-factors with the inclusion of higher order terms decrease, thereby implying a considerable amount of improvement in the rate of convergence.
%
% *********** Fig.5 **********
%
\section{Conclusions}
\label{Conclude}

To summarize, we present threshold enhanced N$^3$LO QCD correction to rapidity distribution of the Higgs boson produced through bottom quark annihilation at the LHC. We show in detail the infra-red structure of the QCD amplitudes at NLO level as well as the cancellation of the various soft and collinear singularities through the summation of all possible degenerate states and the renormalization of the PDFs in order to demonstrate a general framework to obtain threshold corrections to rapidity distributions to all orders in perturbation theory. We have used factorization properties, along with Sudakov resummation of soft gluons and renormalization group invariance to achieve this. The recent result on three loop form factor by Gehrmann and Kara~\cite{Gehrmann:2014vha} and the universal soft distribution obtained in ~\cite{Ahmed:2014cla} provide the last missing information to obtain threshold correction to N$^3$LO for the rapidity distribution of Higgs boson in bottom quark annihilation.  We find the dominance of the threshold contribution over the entire rapidity range at NLO.  We extend this approximation beyond NLO to make predictions for center of mass energies 8 and 13 TeV. We observe that the inclusion of N$^3$LO contributions reduces the scale dependency further, as expected, through the variation of the renormalization and factorization scales around their central values and that K-factors show stability at higher orders.

\section*{Acknowledgement}
 The work of T.A., M.K.M. and N.R. has been partially supported by funding from Regional Center for Accelerator-based Particle Physics (RECAPP), Department of Atomic Energy, Govt. of India. V.R. would like to thank T. Gehrmann for useful discussion.

\bibliography{bbhrmac}
\bibliographystyle{utphysM}
  
\end{document}